# Auxetic Response in Two-Dimensional MXenes with Atomically Defined Perforations

Hossein Darban[1]

*Institute of Fundamental Technological Research, Polish Academy of Sciences, Pawińskiego 5B, 02-106 Warsaw, Poland*

## Abstract

Recent advances in nanoscale fabrication enable atomic-scale manipulation of two-dimensional (2D) materials by introducing engineered pores and perforations. This provides new opportunities to tailor functional properties of 2D materials for applications such as selective ion transport, desalination membranes, and molecular filtration. Despite this progress, the auxetic mechanical behavior of perforated 2D materials has received little attention. In this work, large-scale reactive molecular dynamics (MD) simulations, validated against experimental measurements and first-principles calculations, are employed to investigate the mechanical response of perforated monolayer titanium-based MXene metamaterials. Architectures containing rectangular perforations with straight ligaments and sinusoidally curved ligaments are systematically examined under uniaxial tension and compression over a range of geometric parameters and temperatures, from the onset of deformation to fracture. The results demonstrate that MXene metamaterials exhibit a tunable negative Poisson's ratio (NPR), which can be controlled through the perforation geometry and surface termination. Atomistic stress analysis reveals alternating in-plane shear stresses at the junctions that induce rotational deformation of the ligaments. This rotating-junction mechanism is coupled with out-of-plane deflections arising from the low bending rigidity of atomically thin materials, producing complex three-dimensional deformations. Comparison with graphene metamaterials indicates that the perforation geometry governs qualitative auxetic trends, whereas intrinsic material properties determine quantitative responses. These findings identify MXenes as a versatile candidate for the design of tunable 2D mechanical metamaterials and provide atomistic insight into the interplay between geometry, bending rigidity, and auxetic deformation mechanisms.

**Keywords:** MXenes; Negative Poisson's ratio; Mechanical metamaterials; Perforated two-dimensional materials; Molecular dynamics.

---

[1]- ***Corresponding author:*** *Hossein Darban*; ***E-mail address:*** [hdarban@ippt.pan.pl](hdarban@ippt.pan.pl); ***Tel.:*** *(+48) 22 826 12 81*

# 1. Introduction

The ever-growing human drive to develop new technologies places the discovery, development, and engineering of new materials with unique properties at the forefront of scientific research. One notable example is the emergence of metamaterials exhibiting auxetic behavior, characterized by a Negative Poisson's Ratio (NPR). The first bulk material displaying an NPR was reported in 1987 [1], where a novel foam structure was shown to expand laterally under tensile loading. While the NPR can arise as an intrinsic material property, that is, without deliberate modification, such behavior is generally restricted to specific scenarios. For example, single crystals may exhibit NPR along certain orientations [2]. A prominent crystalline material exhibiting intrinsic NPR is α-cristobalite ($SiO_2$), which was identified in 1992 using laser Brillouin spectroscopy [3]. Auxetic behavior has also been observed in biological systems, including cow teat skin [4]. However, most auxetic materials reported to date rely on engineered microstructural or compositional modifications. In many cases, the NPR originates from specific geometric motifs, such as re-entrant structures, chiral architectures, perforated configurations, and rotating rigid-body mechanisms, as recently reviewed in Refs. [5,6].

On another front of materials development, atomically thin two-dimensional (2D) materials have opened new avenues for engineering materials from the ground up. Like bulk materials, 2D materials can also exhibit auxetic behavior [7]. For instance, single-layer black phosphorus exhibits an intrinsic NPR, arising from its puckered atomic structure [8]. The pucker can be interpreted as a re-entrant geometry composed of two coupled orthogonal hinges. As a consequence of this structural motif, an NPR emerges in the out-of-plane direction when uniaxial deformation is applied parallel to the puckering direction. Other 2D materials, such as graphene, which do not intrinsically exhibit an NPR in their pristine state, can be rendered auxetic through kirigami by introducing patterned cuts. Atomistic simulations first conceptualized this [9] and subsequently demonstrated experimentally using optical lithography to pattern the graphene [10]. Subsequent atomistic simulations predicted that the mechanical response of graphene can be tuned through periodic perforation patterns with rectangular, rhomboidal, elliptical, Cassini, or sinusoidal geometries [11–14]. Auxetic 2D materials can enhance the mechanical performance of composite systems when used as reinforcements, leading to improved fracture toughness, greater shear resistance, and increased energy absorption capacity. In addition, their unusual deformation behavior makes them promising candidates for use in the fabrication of highly sensitive sensors.

Manipulation of 2D materials at the atomic scale has become an active area of research. One important branch of this field focuses on defect and perforation engineering, in which the properties of 2D materials are tailored or entirely new functionalities are introduced through the deliberate creation of defects or pores. Such modifications can be achieved through controlled electron or ion beam irradiation [15–19]. In these approaches, highly energetic particles, such as electrons or ions, interact with the 2D lattice and induce atomic excitations that lead to the formation of defects and pores. The type and size of these defects, ranging from single atomic vacancies to nanoscale pores and perforations, can be controlled by adjusting the energy and nature of the incident particles. For example, sub-angstrom precision in fabricating MXenes with atomically defined pores has been achieved using aberration-corrected scanning transmission electron microscopy (STEM) [20].

Among the most recently discovered members of the nanomaterials family are MXenes, a large class of 2D transition metal carbides, nitrides, and carbonitrides first reported in 2011 [21]. MXenes have the general formula $M_{n+1}X_n T_x$, where M is an early transition metal, X represents carbon or nitrogen, and $T_x$ denotes surface terminations such as –OH, –O, or –F groups. These materials exhibit exceptional optoelectronic, electrochemical, hydrophilic, and mechanical properties [22–28], making them attractive for a wide range of applications, including energy storage and conversion, catalysis, water purification and desalination, electromagnetic interference shielding, communication and optical devices, sensing and actuation, structural composites, and biomedical technologies [21,23–26]. MXenes do not intrinsically exhibit an NPR in their pristine state. Additionally, the auxetic behavior of perforated MXene monolayers has not yet been predicted through atomistic simulations or realized experimentally, analogous to graphene kirigami produced via nanofabrication techniques such as optical lithography [10] or focused ion beam [29]. Recent studies in this area have primarily focused on macroscale MXene-based materials and structures. For example, an MXene-coated acetate substrate with kirigami-inspired cut patterns was investigated to demonstrate the strain-dependent tunability of microwave performance [30]. Other reports [31–33] describe MXene-based foams, hydrogels, and cellular structures that display an NPR.

The present study addresses this scientific gap by providing novel atomistic predictions and mechanistic insights into the auxetic response of perforated 2D MXene monolayers. Large-scale reactive molecular dynamics simulations are performed using the well-established ReaxFF force field [34] to model two titanium carbide–based MXenes, namely $Ti_2CT_x$ and $Ti_3C_2T_x$. Two perforation geometries with straight and curved (sinusoidal) ligaments are considered, while other geometries can be examined using the same modeling framework. Detailed analyses of the influence of perforation geometry on the tensile and compressive responses of these MXene metamaterials are presented. In addition, the underlying deformation mechanisms, stress distributions, and the effects of oxygen surface termination and temperature on the mechanical response are systematically investigated.

The atomistic insights reported here are expected to motivate future experimental investigations of comparable MXene metamaterials fabricated using nanofabrication techniques and characterized by high-resolution microscopy, analogous to the graphene kirigami study reported in Ref. [10] and the in situ tensile testing of single-layer MXenes described in Ref. [29].

## 2. Molecular Dynamics Simulations

We conduct reactive molecular dynamics (MD) simulations using the Large Scale Atomic/Molecular Massively Parallel Simulator (LAMMPS) [35–37] and the ReaxFF force field developed in [34], which is particularly well suited to atomistic modeling of the mechanical behavior of titanium carbide–based MXenes (see Section 3.1). The force field has been applied to investigate the influence of surface terminations, defects, frictional contact, indentation, and fracture processes on the mechanical behavior of $Ti_3C_2O_2$ and $Ti_2CO_2$ MXenes [38–43]. The initial atomic configurations of the $Ti_3C_2O_2$ and $Ti_2CO_2$ MXene metamaterials with straight and curved ligaments are shown in Fig. 1(a) and (b). In the curved design, each ligament follows a sinusoidal profile, reaching its maximum amplitude at the center of the unit cell and its minimum at the edges. Each configuration consists of five square unit cells, each with dimensions $30 \times 30$ nm² in the $x$- and y-directions. The geometric features of the individual unit cells are illustrated in Fig. 1(c) and (d). All initial configurations and visualizations were generated using the Open Visualization Tool (OVITO) [44].

The unit cell size is deliberately chosen to be approximately two orders of magnitude larger than the characteristic bond lengths in MXenes. This scale separation ensures that the mechanical response is governed primarily by structural geometry rather than atomic discreteness, thereby minimizing artificial size effects arising from the ratio of perforation dimensions to interatomic bond lengths. Such a choice supports the validity of interpreting the results within a continuum mechanics framework, despite the atomistic resolution of the simulations.

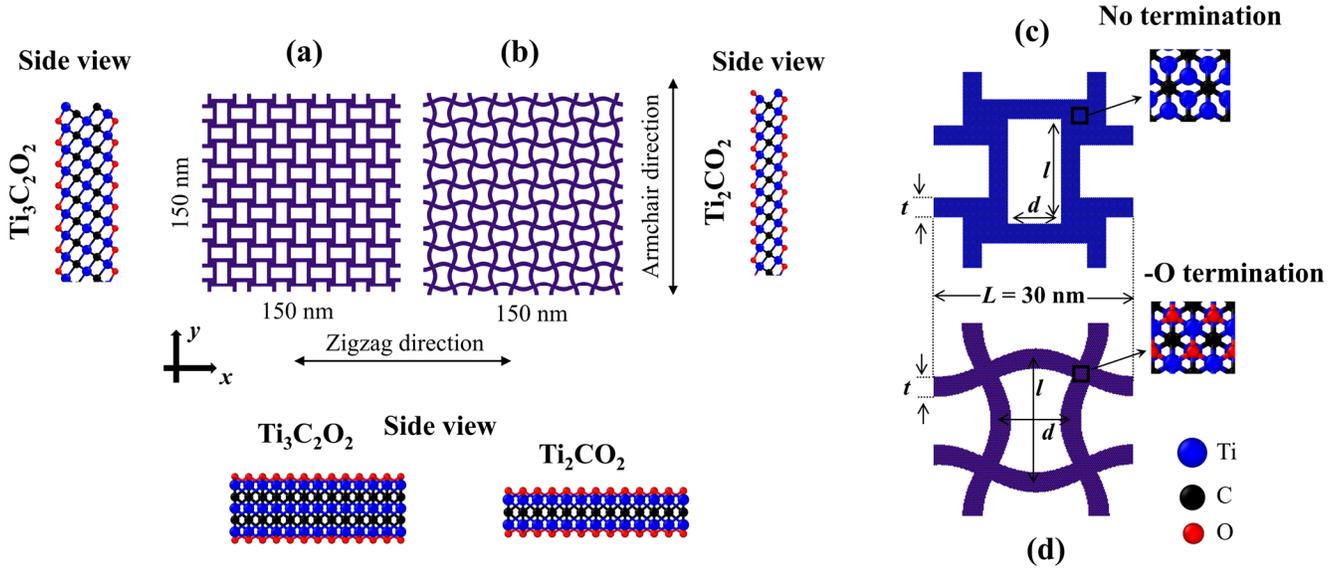

**Fig. 1** (a) and (b) Initial atomic configurations of Ti$_3$C$_2$O$_2$ and Ti$_2$CO$_2$ MXene metamaterials with straight and curved ligaments. The top view and part of the side view are presented. (c) and (d) Exemplary representative unit cells.

Three-dimensional (3D) simulations are performed with periodic boundary conditions applied in the in-plane directions to represent a large MXene monolayer. Along the out-of-plane ($z$) direction, a periodic boundary condition is also imposed, with a 6 nm-thick vacuum layer, which effectively corresponds to a non-periodic boundary with sufficient space to accommodate out-of-plane deformations. A timestep of 0.2 femtoseconds (fs) is assumed for all the simulations. To eliminate residual stresses in the initial configurations, energy minimization was first performed using the conjugate gradient (CG) algorithm with the Polak–Ribiere formulation, employing energy and force tolerances of $10^{-10}$ (dimensionless) and $10^{-10}$ (kcal/mol)/Å, respectively. The velocity form of the Stoermer-Verlet time integration algorithm (velocity-Verlet) is used [45].

Following energy minimization, atomic velocities are assigned according to the target temperatures of 1 K, 150 K, 300 K, and 450 K. Before mechanical loading, the system is equilibrated for 100 picoseconds (ps) using the NPT ensemble at the prescribed temperature, and an in-plane pressure of 0 Pa. Temperature and pressure are regulated using the Nosé–Hoover thermostat and barostat. During equilibration, thermodynamic quantities, including temperature, total energy, and forces, are monitored to ensure that the system reaches thermal and structural equilibrium. After equilibration, the MXene monolayers are deformed along the zigzag direction (x-axis) at a constant engineering strain rate of 0.001 ps$^{-1}$ using the NPT ensemble. At this stage, the pressure along the $y$-direction was maintained at 0 Pa to ensure a uniaxial stress condition. Both tensile and compressive loadings are considered. The axial and lateral strains are determined from the change in the monolayer's dimensions relative to the initial lengths. The Poisson's ratio is defined as the negative ratio of the lateral strain to the axial strain, $-\varepsilon_y/\varepsilon_x$.

During deformation, the axial stress is computed using the virial formulation. To convert the virial quantities reported by LAMMPS, which have units of stress multiplied by volume, into true stress values, an effective volume was assigned to the perforated MXene monolayer. First, the volume of the pristine (non-perforated) monolayer was estimated as $V_0=L_xL_yL_z$, where $L_x$ and $L_y$ are the in-plane dimensions of the sheet (150 nm), and $L_z$ is the thickness of the MXene defined as the distance between the uppermost and lowermost atomic layers (see Table 1 for the values of thicknesses). This reference volume was divided by the total number of atoms in the pristine MXene monolayer to obtain an average atomic volume. Following the introduction of perforations, the remaining number of atoms in the MXene monolayer was counted, and the effective volume of the perforated structure was computed by multiplying this atom count by the previously determined average atomic volume. This normalization procedure provides a consistent and physically meaningful approximation of the effective volume for stress calculation, particularly for large systems considered in this study with 0.2-1.3 million atoms.

All computations were performed on the Helios supercomputer, hosted by the Academic Computer Centre CYFRONET AGH in Kraków, Poland. Between one and four computing nodes were employed, each equipped with 192 cores provided by dual AMD EPYC 9654 96-core processors (2.4 GHz) and 384 GB of RAM. The total wall-clock time for each simulation depended on the system size and simulation duration. For instance, the $Ti_3C_2O_2$ simulations presented in Section 3.6 required approximately 120 hours to complete on four computing nodes.

## 3. Results and Discussions

### 3.1. Simulation validation

Before presenting the results for MXene metamaterials, the accuracy of the ReaxFF force field developed in Ref. [34] and employed in this study was validated. The validation considers all three MXenes investigated in this work. Both structural and mechanical properties were examined, including the lattice parameter, monolayer thickness, bond lengths, and the directional Young modulus. The MD results were compared with available first-principles density functional theory (DFT) calculations and experimental data reported in the literature. Numerous DFT studies on these MXenes report very similar structural and mechanical properties [46–50]. In the present work, comparisons are made only with representative results reported in the literature [47,51], and the obtained MD values are consistent with other DFT data. In addition, the MD predictions were compared with experimental measurements available for $Ti_2CT_x$ and $Ti_3C_2T_x$ MXenes. Details of the simulation procedures used to determine the MD results are provided in Supplementary Material S1. The comparison between MD, DFT, and experimental data is summarized in Table 1.

**Table 1:** Comparison of in-plane lattice parameter *a*, monolayer thickness, bond lengths, and elastic constants obtained from present MD simulations using the ReaxFF force field developed in [34] with reported density functional theory (DFT) calculations and experimental measurements for the studied MXenes.

| Property | $Ti_2C$, present MD | $Ti_2C$ DFT | $Ti_2CO_2$, present MD | $Ti_2CO_2$ DFT | $Ti_2CT_x$ Exp. | $Ti_3C_2O_2$, present MD | $Ti_3C_2O_2$ DFT | $Ti_3C_2T_x$ Exp. |
|---|---|---|---|---|---|---|---|---|
| *a* (Å) | 3.062 | 3.035 [51] | 3.040 | 3.033 [51] | 3.001 [52] | 3.059 | 3.039 [51] | 3.051 [53] |
| Thickness (Å) | 2.292 | 2.305 [51] | 4.454 | 4.431 [51] | 6.700 [a] [54] | 6.921 | 6.959 [51] | 9.400 [a] [54] |
| $Ti^{upper}$–C bond length (Å) | 2.105 | 2.097 [51] | 2.255 | 2.185 [51] | 2.110 [52] | 2.185 | 2.193 [51] | 2.209 [52] |
| $Ti^{middle}$–C bond length (Å) | – | – | – | – | – | 2.155 | 2.158 [51] | 2.101 [52] |
| Ti–O bond length (Å) | – | – | 1.915 | 1.973 [51] | 1.909 [52] | 1.935 | 1.976 [51] | 1.972 [52] |
| Young's modulus (GPa), zigzag direction | 677 | 620 [47] | 459 | 593 [47] | - | 479 | 468[b] [47] | 330[c] [55] 484[d] [29] |
| Young's modulus (GPa), armchair direction | 677 | 600 [47] | 469 | 540 [47] | - | 486 | 491[b] [47] | 330[c] [55] 484[d] [29] |

a: The value is reported for MXene with OH terminations, which has a higher thickness than O-terminated MXene.

b: Data obtained from a curve.

c: From nanoindentation of a $Ti_3C_2T_x$ MXene membrane, no directional data were presented.

d: From the direct tension of a $Ti_3C_2T_x$ MXene membrane, no directional data were presented.

For all three MXenes, the in-plane lattice parameter predicted by MD simulations differs by approximately 1% from the DFT and experimental values. Similarly, the monolayer thickness obtained from MD simulations agrees with DFT results within 1% error. The experimental monolayer thickness of $Ti_2CT_x$ and $Ti_3C_2T_x$, determined from X-ray diffraction (XRD) measurements in [54], is also in good agreement with values reported from atomically resolved transmission electron microscopy (TEM) [56]; however, these experimental measurements correspond to –OH-terminated MXenes. Consequently, the experimental thickness values listed in Table 1 are larger than the MD and DFT predictions for –O terminated MXenes, which can be attributed to the additional O–H bond length. Additionally, the bond lengths predicted by the MD simulations are in good agreement with those obtained from DFT and experiments.

Two experimental measurements of the Young's modulus for $Ti_3C_2T_x$ MXenes are available. In the first study [55], nanoindentation tests using atomic force microscopy on single-layer $Ti_3C_2T_x$ yielded a Young's modulus of $330 \pm 30$ GPa. More recently, advances in nanoscale sample preparation and in-situ tensile testing within an electron microscope reported a Young's modulus of $484 \pm 13$ GPa for single-layer $Ti_3C_2T_x$ [29]. Our MD predictions are in excellent agreement with these experimental results, particularly the direct tensile measurements, and also closely match the DFT predictions. These validations demonstrate that the ReaxFF force field employed here reliably reproduces both structural and mechanical properties of the studied MXenes.

### 3.2. $Ti_2CT_x$ MXene metamaterials with straight ligaments

The stress-strain, axial strain-lateral strain, and axial strain-Poisson's ratio curves related to the $Ti_2C$ (without termination) and $Ti_2CO_2$ MXene metamaterial with straight ligaments (rectangular perforation), subjected to a uniaxial tension along the zigzag direction, are illustrated in Fig. 2 and Fig. 3. Results are shown for a temperature of 1 K and different dimensionless geometric parameters, namely the ligament thickness to unit cell length ($t/L$) and rectangular perforation aspect ratios ($l/d$).

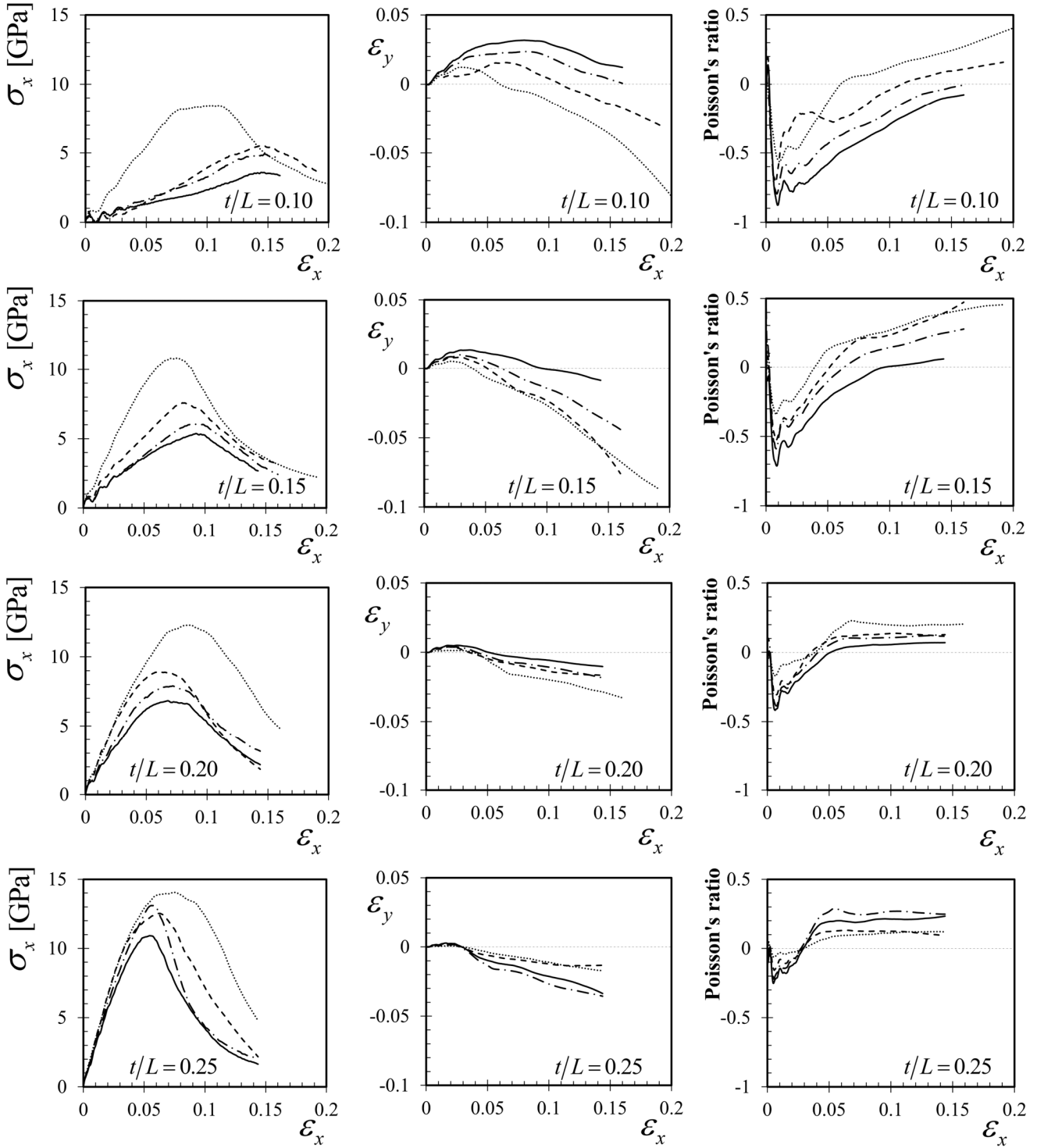

**Fig. 2** Stress–strain response, axial and lateral strain evolution, and Poisson's ratio curves of the Ti$_2$C MXene metamaterial with straight ligaments subjected to uniaxial tension along the zigzag direction. Results are presented at 1 K for different geometric parameters.

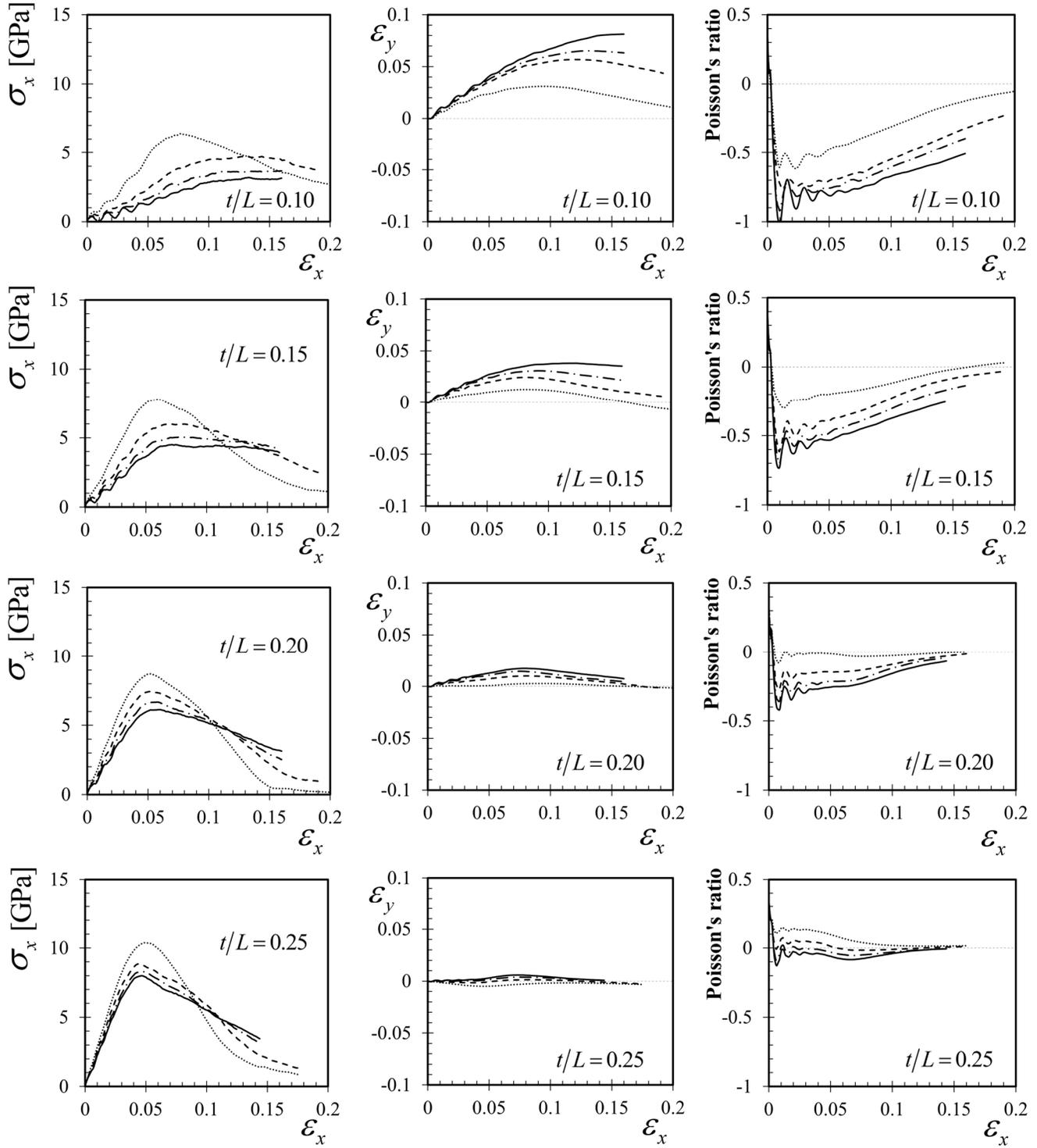

**Fig. 3** Stress–strain response, axial and lateral strain evolution, and Poisson's ratio curves of the $Ti_2CO_2$ MXene metamaterial with straight ligaments subjected to uniaxial tension along the zigzag direction. Results are presented at 1 K for different geometric parameters.

The key observation is that both Ti$_2$C and Ti$_2$CO$_2$ MXene metamaterials exhibit a negative Poisson's ratio (NPR). This auxetic behavior originates primarily from the "rotating rigid units" deformation mechanism [57], in which the small squares located at the corners of each unit cell (see Fig. 1(c)) rotate relative to one another under tensile loading. This deformation mode is accompanied by the out-of-plane deformations as discussed and illustrated in Section 3.4. The negative Poisson's ratio initially intensifies as the lateral strain approaches its maximum, after which the lateral strain begins to decrease, causing the negative Poisson's ratio to diminish and eventually become positive at higher strains. This general trend is consistent across all cases presented in Fig. 2 and Fig. 3.

The dimensionless geometric parameters play a critical role in governing the mechanical response of the MXene metamaterials. For a constant ligament thickness to unit cell length ratio ($t/L$), increasing the rectangular perforation aspect ratio significantly reduces both the initial stiffness and the ultimate tensile strength of the MXene metamaterials. The reduction in strength is primarily attributed to the higher stress concentrations induced by perforations with larger aspect ratios, which promote earlier failure. The lower initial stiffness observed at higher aspect ratios arises because, for a fixed ligament thickness, the ligaments become longer in the loading direction, resulting in a more compliant response. In contrast, for a fixed perforation aspect ratio, increasing the ligament thickness results in a stiffer mechanical response and a higher ultimate strength due to enhanced load-carrying capacity of the ligaments.

The geometric parameters also strongly influence the auxetic response of the MXene metamaterials. At a given ligament thickness, a higher perforation aspect ratio leads to a more pronounced NPR. However, for a fixed perforation aspect ratio, increasing the ligament thickness weakens auxetic behavior by reducing the dominant deformation mechanism associated with rotating rigid units.

In addition, the -O surface termination reduces both the stiffness and the ultimate strength of the MXene metamaterials. This softening effect of the oxygen surface termination is in agreement with the first-principles predictions reported in [47]. This mechanically softer response alters the mechanisms of ligament deformation and, consequently, the auxetic behavior. In several cases, Ti$_2$CO$_2$ MXene metamaterials exhibit an NPR over a wider strain range compared to their non-terminated counterparts. However, this trend is not universal. For example, the Ti$_2$CO$_2$ MXene metamaterial with $l/d=2$ and $t/L=0.25$ does not exhibit an NPR, whereas the corresponding Ti$_2$C configuration shows a slight auxetic response at lower strains. Interestingly, for $l/d=2$ and $t/L=0.20$, the Ti$_2$CO$_2$ MXene metamaterial exhibits an almost-zero Poisson's ratio over most of the deformation process.

Results presented in Fig. 2 and Fig. 3 demonstrate that the mechanical response of MXene metamaterials can be systematically tailored to induce auxetic behavior, to control the magnitude of the auxetic response, or even to trigger a transition between auxetic and non-auxetic regimes through the combined effects of geometric design and surface termination.

The results for the same $Ti_2C$ (without termination) and $Ti_2CO_2$ MXene metamaterial with straight ligaments (rectangular perforation), this time subjected to a uniaxial compression along the zigzag direction, are illustrated in Fig. 4 and Fig. 5. The general trends in the results for MXene metamaterials under compression are similar to those observed above for tensile loading. Both MXene metamaterials shrink laterally under compression, indicating an NPR. In contrast to the tensile responses shown in Fig. 2 and Fig. 3., where the post-peak stress reduction results from crack initiation and propagation, the stress decline under compression is primarily due to ligament buckling. This also explains why the $Ti_2CO_2$ metamaterial exhibits a higher maximum compressive stress compared with its non-terminated counterpart. The presence of –O surface terminations effectively increases the out-of-plane thickness of the MXene sheet, which enhances the bending rigidity of the ligaments and consequently raises the buckling stress [58]. Furthermore, in the compression regime, the auxetic response arises from two coupled deformation mechanisms: the "rotating rigid units" mechanism and out-of-plane deformations, which will be discussed in Section 3.4. The influence of geometric parameters on the stress–strain response, as well as on the evolution of axial and lateral strains and Poisson's ratio, generally follows trends similar to those observed under tensile loading.

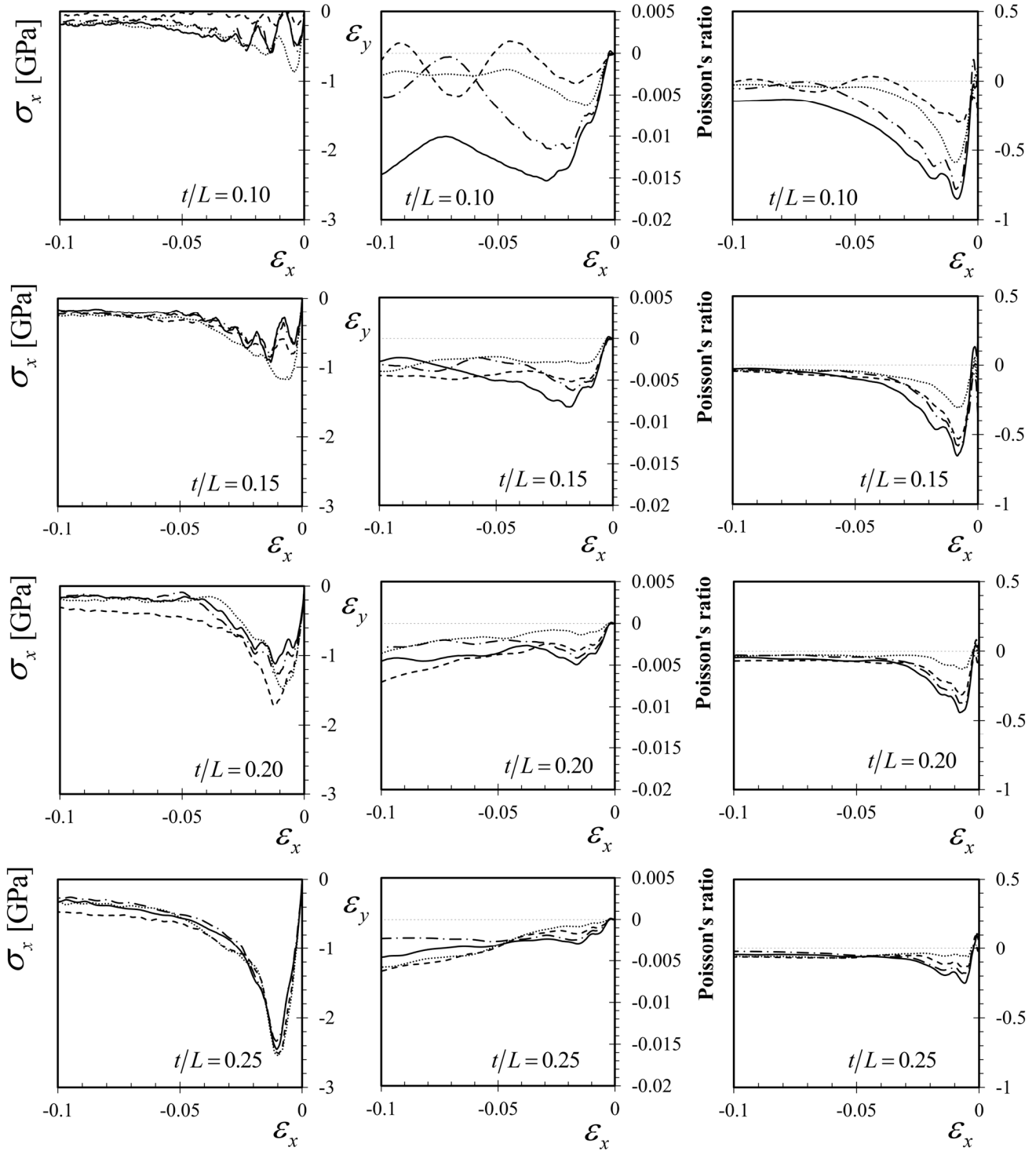

**Fig. 4** Stress–strain response, axial and lateral strain evolution, and Poisson's ratio curves of the Ti$_2$C MXene metamaterial with straight ligaments subjected to uniaxial compression along the zigzag direction. Results are presented at 1 K for different geometric parameters.

**Ti$_2$CO$_2$ under compression** 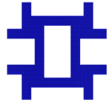 ······ $l/d = 2$   --- $l/d = 3$   -·-· $l/d = 4$   —— $l/d = 5$

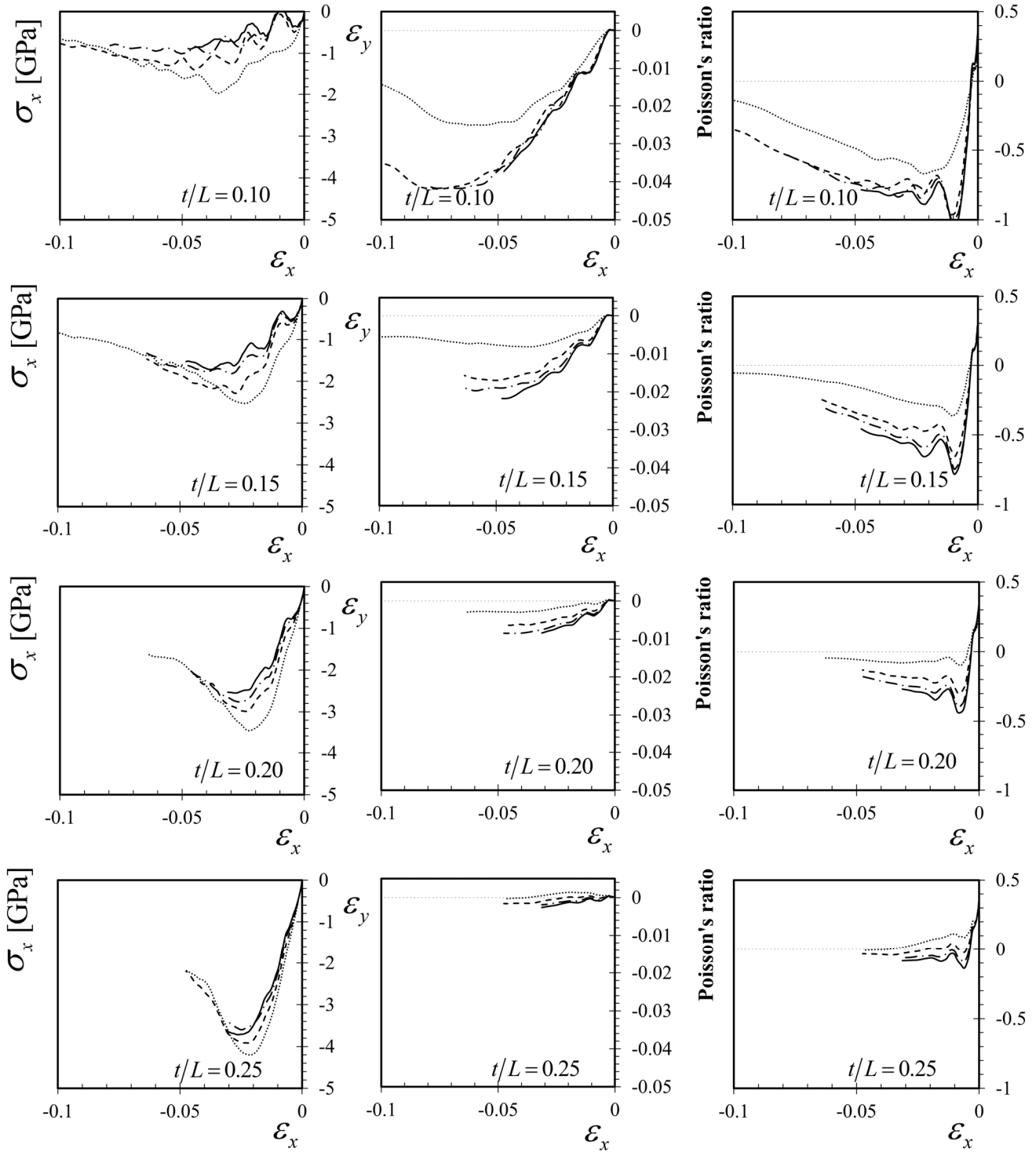

**Fig. 5** Stress–strain response, axial and lateral strain evolution, and Poisson's ratio curves of the Ti$_2$CO$_2$ MXene metamaterial with straight ligaments subjected to uniaxial compression along the zigzag direction. Results are presented at 1 K for different geometric parameters.

### 3.3. Ti$_2$CT$_x$ MXene metamaterials with curved ligaments

The stress-strain, axial strain-lateral strain, and axial strain-Poisson's ratio curves related to the Ti$_2$C (without termination) and Ti$_2$CO$_2$ MXene metamaterial, with curved (sinusoidal) ligaments, subjected to a uniaxial tension along the zigzag direction, are illustrated in Fig. 6 and Fig. 7. Results are shown for simulations at the temperature of 1 K and for different dimensionless geometric parameters, namely the ligament thickness to unit cell length ($t/L$) and perforation aspect ratios ($l/d$).

Comparing the stress–strain curves of MXene metamaterials with sinusoidally curved ligaments to those with straight ligaments examined in the previous section reveals a fundamentally different mechanical response. In the former case, the stress–strain behavior exhibits a characteristic J-shaped profile, which has previously been reported for similar macroscale bioinspired hierarchical lattice materials [59,60], soft network composites [61], stretchable electronics [62], and graphene-based kirigami metamaterials with sinusoidal geometries [11]. The initial portion of the stress–strain curves is characterized by a compliant response dominated by bending of the curved ligaments due to rotations at their junctions. In this regime, the stiffness is primarily governed by the bending rigidity of the ligaments, which depends on the geometric parameters such as ligament thickness and perforation aspect ratio. Beyond a critical strain, the deformation mechanism transitions to a tension-dominated regime, in which stiffness is primarily governed by the intrinsic stiffness of the MXene lattice. The auxetic response reaches its maximum near this critical strain, after which the lateral strain gradually decreases as the deformation enters the tension-dominated stage. The value of the critical strain depends on both the surface termination and the geometric parameters. For example, in the Ti$_2$C MXene metamaterial with $l/d = 13/7$ and $t/L = 0.05$, the transition in deformation mechanism occurs at an axial strain of approximately 5%, whereas for the Ti$_2$CO$_2$ MXene metamaterial with $l/d = 3$ and $t/L = 0.05$, the critical strain exceeds 10%. It should be noted that throughout the loading process, the deformation mechanisms remain complex due to out-of-plane deformations. These effects will be discussed in more detail in Section 3.4.

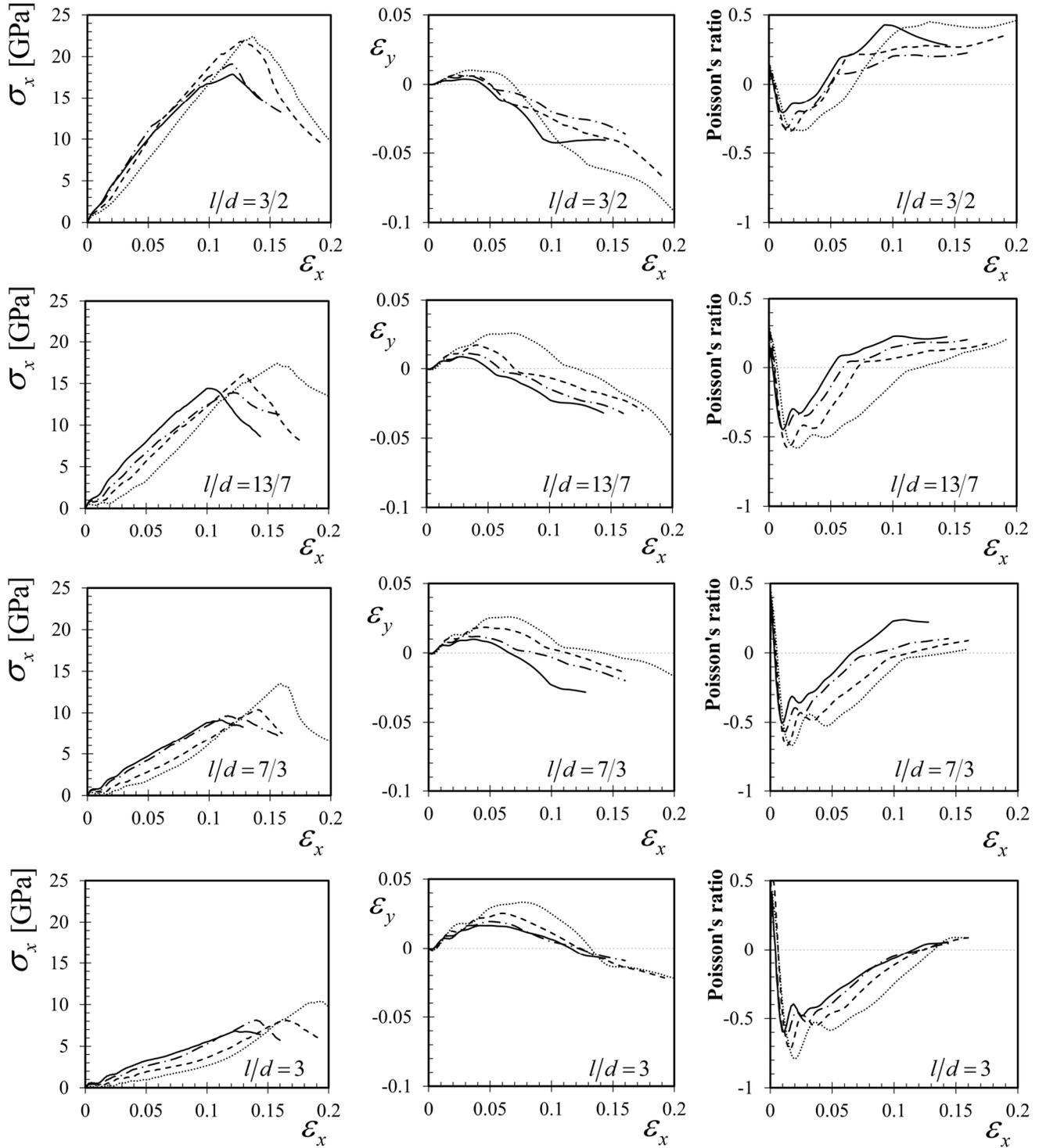

**Fig. 6** Stress–strain response, axial and lateral strain evolution, and Poisson's ratio curves of the Ti$_2$C MXene metamaterial with curved ligaments subjected to uniaxial tension along the zigzag direction. Results are presented at 1 K for different geometric parameters.

**Ti$_2$CO$_2$ under tension** 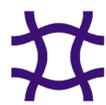   ······ $t/L = 0.05$   - - - $t/L = 0.075$   - · - · $t/L = 0.10$   —— $t/L = 0.125$

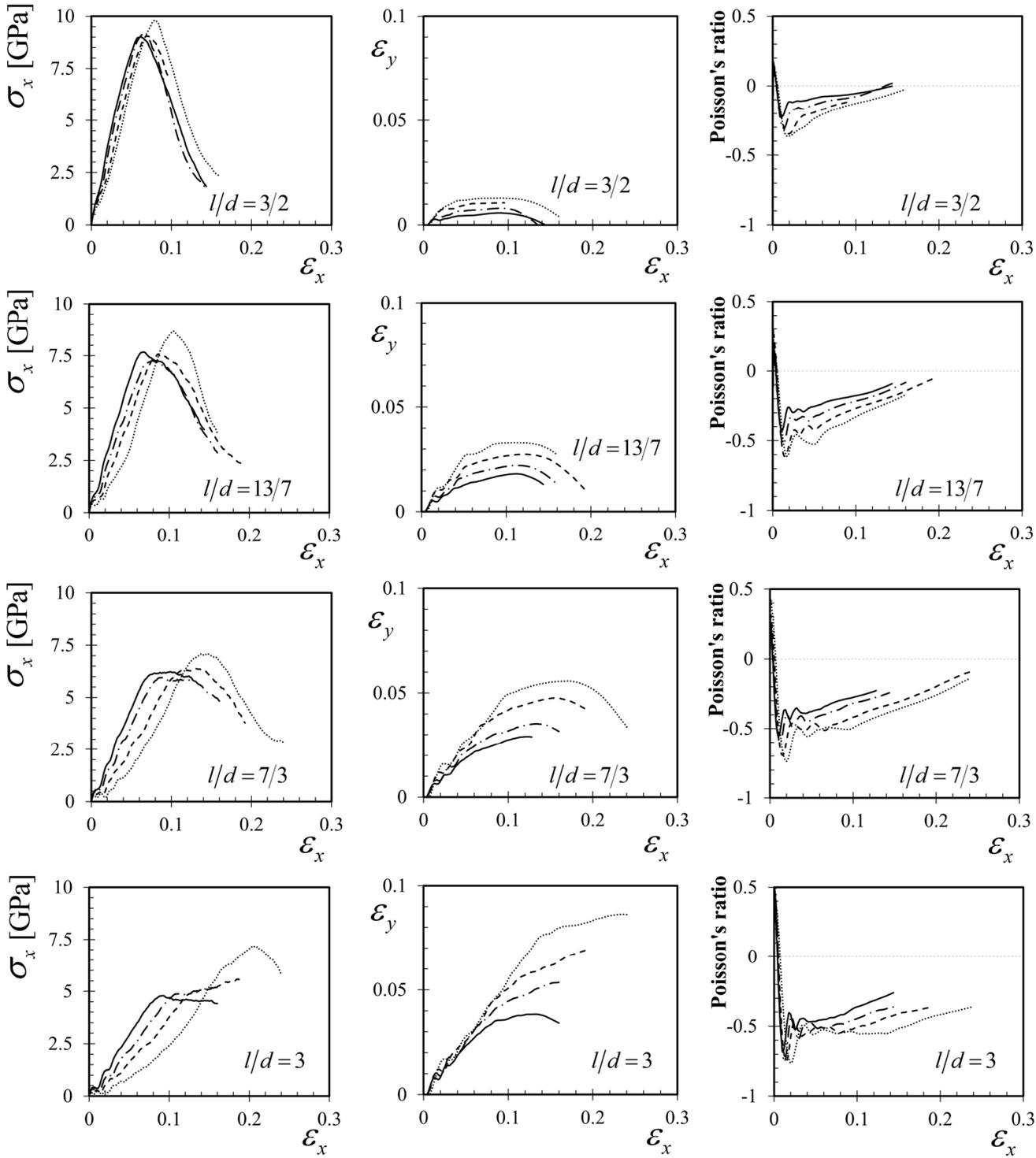

**Fig. 7** Stress–strain response, axial and lateral strain evolution, and Poisson's ratio curves of the Ti$_2$CO$_2$ MXene metamaterial with curved ligaments subjected to uniaxial tension along the zigzag direction. Results are presented at 1 K for different geometric parameters.

The stress–strain response and the magnitude of the negative Poisson's ratio (NPR) are strongly influenced by the geometric parameters of the perforations. For both MXene metamaterials with and without oxygen termination, the highest tensile strength is observed in structures with smaller ligament thickness, consistent with the trend previously observed for MXene metamaterials with rectangular perforations in Fig. 2 and Fig. 3. In general, MXene metamaterials with curved ligaments and thinner ligaments exhibit a softer stress–strain response. This behavior arises because thinner ligaments significantly reduce bending rigidity and facilitate rotation at the junctions. As a result, the deformation remains bending-dominated over a wider strain range. Furthermore, as the perforation aspect ratio increases, resulting in more slender ligaments, the stress–strain response becomes softer, and the influence of ligament thickness on stiffness becomes more pronounced. Higher perforation aspect ratios combined with thinner ligaments also lead to a more pronounced auxetic response. Interestingly, $Ti_2CO_2$ MXene metamaterials tend to exhibit an NPR over a wider strain range compared with their non-terminated counterparts.

Similar results for $Ti_2C$ and $Ti_2CO_2$ MXene metamaterials with sinusoidally curved ligaments subjected to uniaxial compression along the zigzag direction are presented in Fig. 8 and Fig. 9. Unlike the tensile case, the compressive stress–strain curves do not exhibit a J-shaped profile. In compression, the deformation is primarily governed by bending of the curved ligaments and rotations at the junctions, followed by ligament buckling, which leads to the post-peak response characterized by a reduction in axial stress (see Section 3.4 for a more detailed analysis). Consequently, surface terminations and perforation geometries with thicker ligaments and lower aspect ratios increase the maximum compressive stress. In addition, MXene metamaterials with curved ligaments exhibit a NPR under compression due to the combined effects of the "rotating rigid units" mechanism and out-of-plane deformations. Perforation geometries with thinner curved ligaments and higher aspect ratios further intensify the auxetic response.

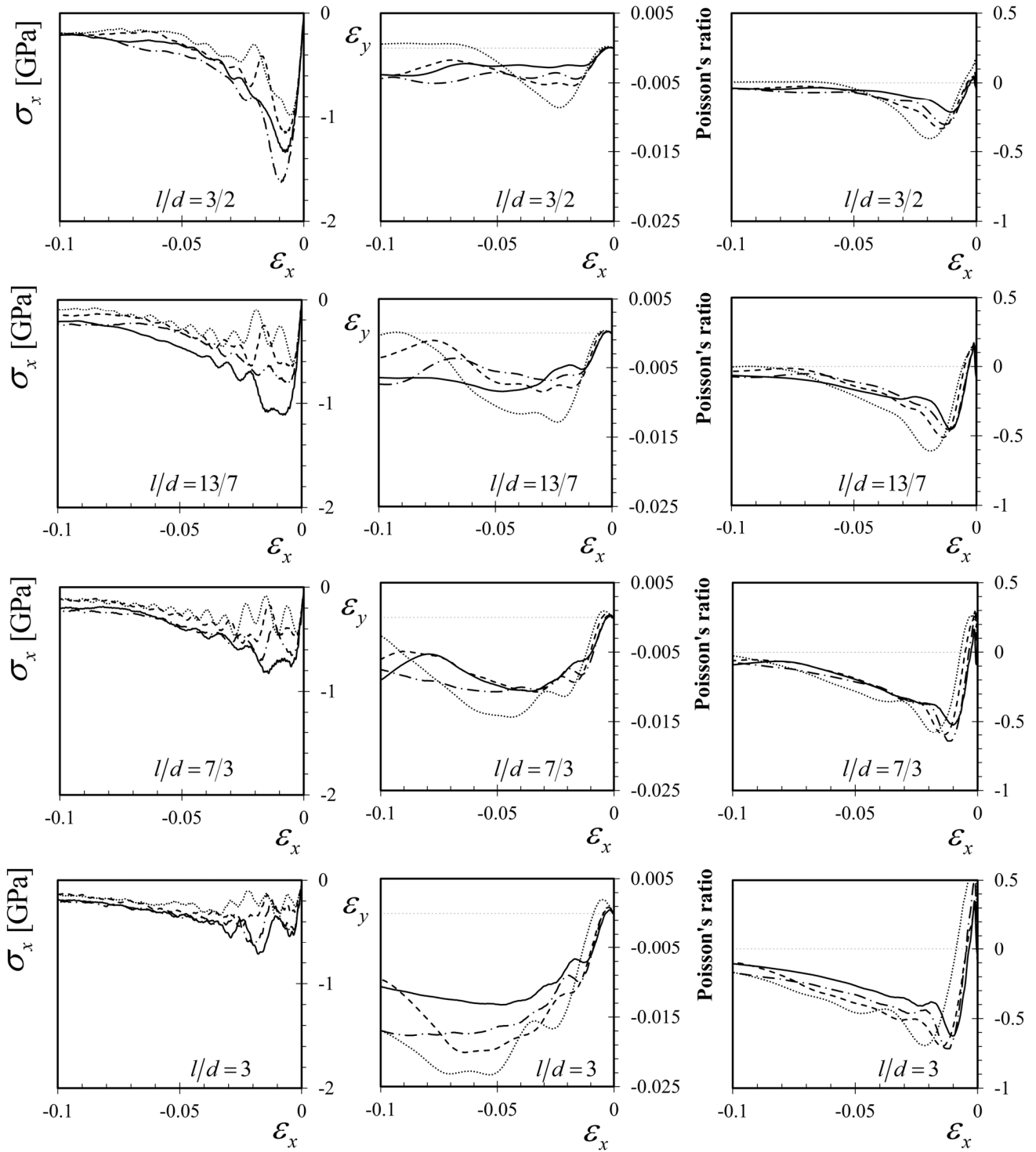

**Fig. 8** Stress–strain response, axial and lateral strain evolution, and Poisson's ratio curves of the $Ti_2C$ MXene metamaterial with curved ligaments subjected to uniaxial compression along the zigzag direction. Results are presented at 1 K for different geometric parameters.

**Ti$_2$CO$_2$ under compression** 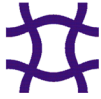 ·······  $t/L=0.05$ --- $t/L=0.075$ -·-· $t/L=0.10$ —— $t/L=0.125$

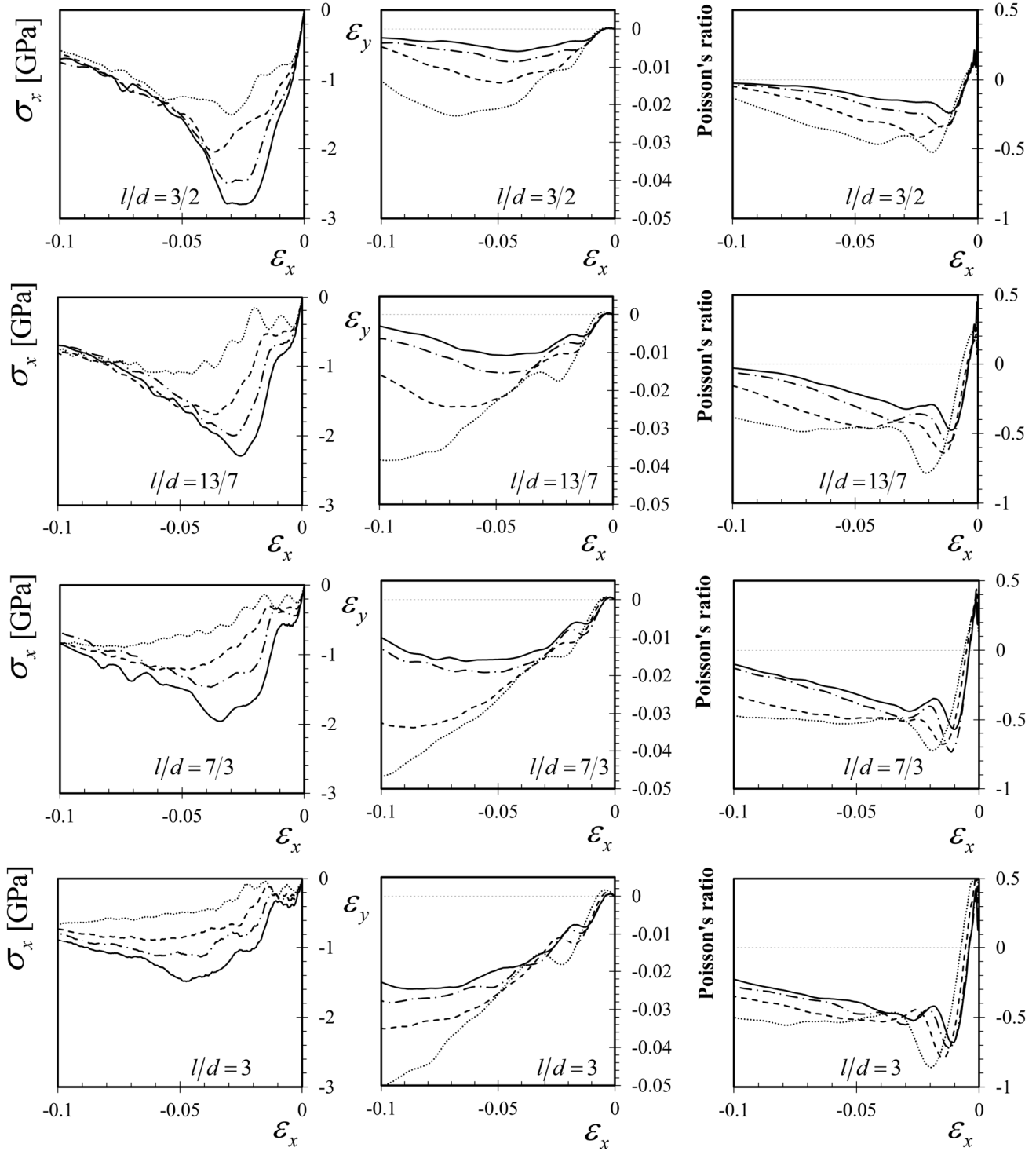

**Fig. 9** Stress–strain response, axial and lateral strain evolution, and Poisson's ratio curves of the Ti$_2$CO$_2$ MXene metamaterial with curved ligaments subjected to uniaxial compression along the zigzag direction. Results are presented at 1 K for different geometric parameters.

## 3.4. Deformation and Stress Analyses in Ti$_2$CT$_x$

It is essential to develop a detailed atomistic-scale understanding of the deformation mechanisms responsible for the auxetic behavior in MXene metamaterials, as well as the associated stress distributions. Such insights are crucial for the rational design of MXene architectures with tailored mechanical responses. In this section, we present detailed analyses of these mechanisms and the corresponding stress fields. The top and side views (x–y and y–z planes) of the atomic configurations under increasing applied strain, for both tensile and compressive loading of the Ti$_2$C MXene metamaterial with straight ligaments, are shown in Fig. 10.

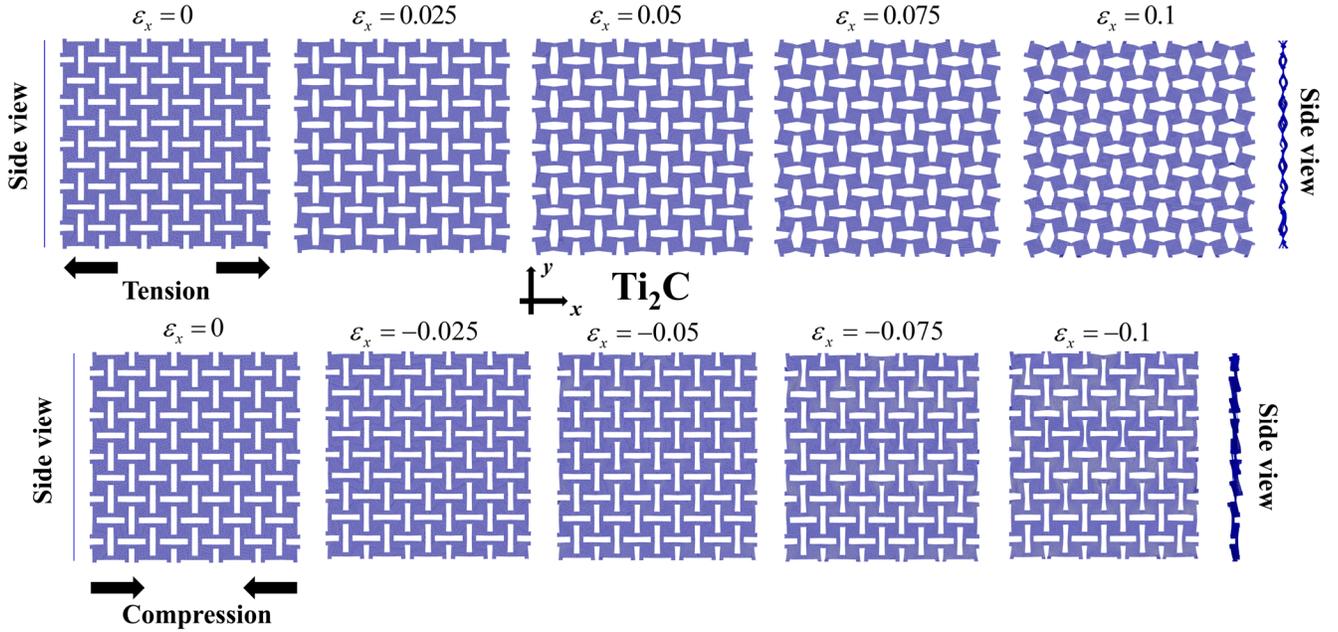

**Fig. 10** Atomic configurations (top view) of the Ti$_2$C MXene metamaterial with straight ligaments subjected to uniaxial tension or compression along the zigzag direction. Results are shown for a temperature of 1 K and $t/L=0.1$, $l/d=5$, and different strain levels. Side views of the initial and last snapshots are also shown.

Under tensile loading, the square units at the ligament junctions rotate, activating the characteristic deformation mechanism of "rotating rigid units". These rotations enlarge the perforations and induce lateral expansion of the MXene metamaterial. As the applied tensile strain increases, the rotations become more pronounced and progressively promote the auxetic response. However, due to the inherently low bending rigidity of 2D materials, noticeable out-of-plane deflections are observed, as evidenced by the side view at an axial strain of 10%. These local out-of-plane deformations intensify with increasing strain and partially suppress lateral expansion, thereby reducing NPR magnitude. This behavior is one of the reasons for the decrease in Poisson's ratio observed after its peak values in Fig. 2.

Under compression, a similar but mechanically reversed "rotating rigid units" mechanism is activated. In this case, the junction rotations lead to contraction of the perforations. Moreover, perforations oriented perpendicular to the loading direction contract more significantly due to the applied compressive stress. Ligament buckling induces local out-of-plane deflections, which, in contrast to the tensile case, can promote the auxetic response. This explains why, in certain cases shown in Fig. 4, the lateral strain becomes increasingly negative at higher compressive strains. These local out-of-plane deflections, caused by ligament buckling under compression, are more pronounced than those observed under tensile loading.

The deformation mechanisms described above are complex and involve localized 3D deformations. Nevertheless, they can be broadly classified into two categories: in-plane deformations mainly characterized by rotations of the junctions, and out-of-plane deformations arising from ligament buckling and the intrinsically low bending rigidity of 2D MXenes. Movies S2 and S3 (tension) and S4 and S5 (compression), provided in the Supplementary Materials, illustrate the complex deformation processes in the $Ti_2C$ MXene metamaterials with straight ligaments.

To further elucidate the dominant deformation mechanisms, the lateral and out-of-plane atomic displacements of the central unit cell in $Ti_2CO_2$ MXene metamaterials with straight ligaments subjected to uniaxial tension along the zigzag direction (x-axis) are presented in Fig. 11. The results are presented as a function of applied strain for three different perforation geometries. From (a) to (b), and from (d) to (e), the ligament thickness is kept constant while the aspect ratio of the rectangular perforation increases. In contrast, from (b) to (c), and from (e) to (f), the perforation aspect ratio remains constant while the ligament thickness increases.

Fig. 11(a), (b), and (c) illustrate the lateral displacement within the unit cell, which is responsible for the NPR. The displacement field is nearly symmetric about the *x*-axis and indicates that the maximum lateral displacement occurs at the corners of the square junctions. As evident from the figures, the lateral displacement primarily results from the rotation of these square junctions. With increasing applied strain, the rotations at the junctions become more pronounced, leading to larger lateral displacements and consequently a stronger auxetic response. These "rotating rigid units" effectively cause the perforations to widen during deformation. For the configuration with thicker ligaments, cracks are visible at the corners of the perforations at an axial strain of approximately 0.06.

The out-of-plane deformations are illustrated in Fig. 11(d), (e), and (f). These deformations exhibit complex distributions within the unit cells and are strongly influenced by the perforation geometry. As shown in the figures, the unit cells exhibit both positive and negative out-of-plane displacements, thereby transforming the initially planar 2D unit cell into a locally distorted 3D configuration. Such intricate out-of-plane deformations, coupled with the in-plane deformation mechanisms, highlight the distinctive 3D deformation characteristics of atomically thin metamaterials. The deformation mechanisms shown in Fig. 11 and discussed here are consistent with those previously reported for graphene metamaterials in Ref. [11,13,14]. Movie S6 provided in the Supplementary Materials illustrates the lateral and out-of-plane atomic deformations during the tensile loading of a $Ti_2CO_2$ MXene metamaterial with straight ligaments.

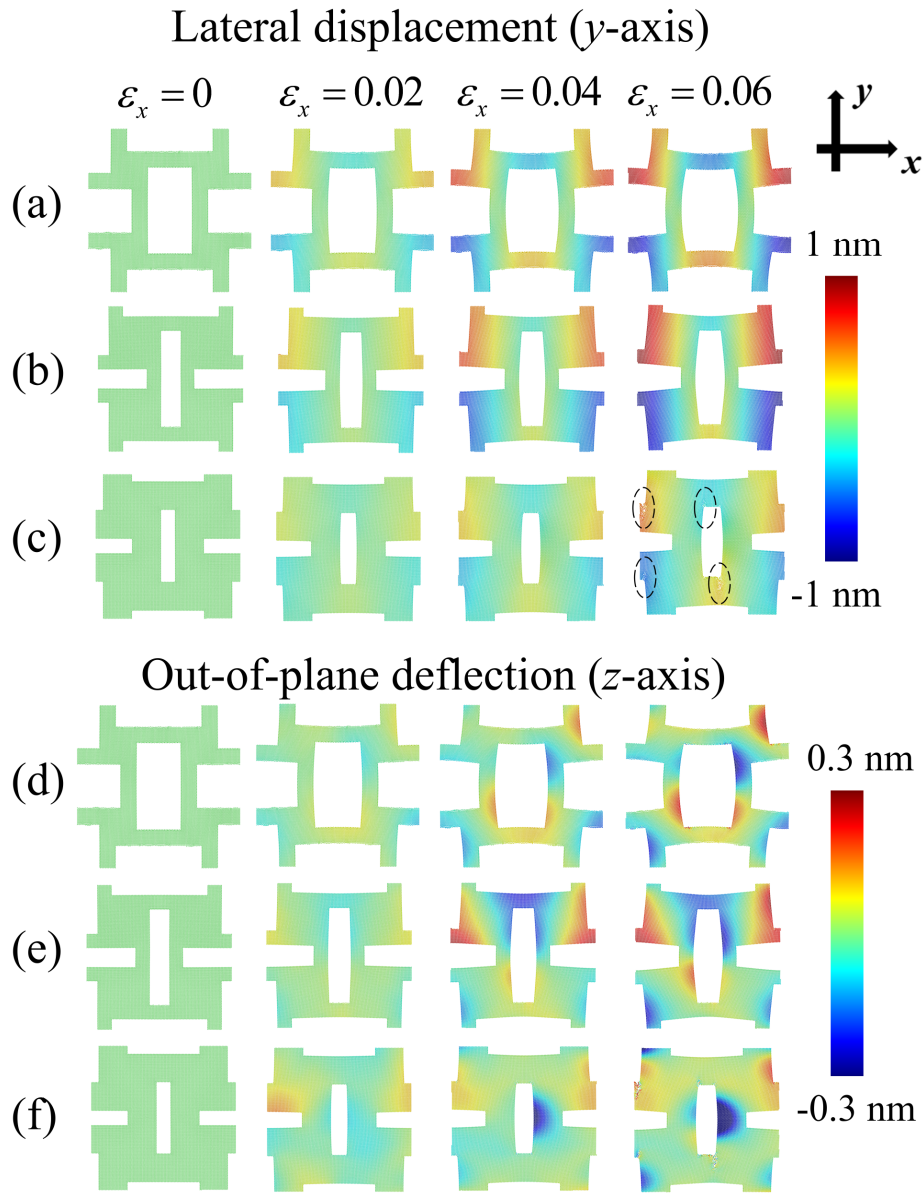

**Fig. 11** Lateral and out-of-plane atomic displacements of the central unit cell of $Ti_2CO_2$ MXene metamaterials with straight ligaments subjected to uniaxial tension along the zigzag direction (x-axis). Results are shown for a temperature of 1 K and $t/L=0.1$ ((a), (b), (d), and (e)), $t/L=0.2$ ((c) and (f)), $l/d=2$ ((a) and (d)), and $l/d=5$ ((b), (c), (e), and (f)), and different strain levels. Cracked and damaged regions are encircled in (c).

The top and side views of the atomic configurations under increasing applied strain, for both tensile and compressive loading of the $Ti_2CO_2$ MXene metamaterial with sinusoidally curved ligaments, are shown in Fig. 12. Similar to the deformation behavior discussed above for MXene metamaterials with straight ligaments, the "rotating rigid units" mechanism is the dominant deformation mode under both tension and compression and is primarily responsible for the auxetic response. As in the case of straight ligaments, this mechanism is coupled with out-of-plane deformations, resulting in a complex 3D deformation pattern. These out-of-plane deflections are clearly visible in the side views at the applied strain of 0.1 and are more pronounced under compression due to ligament buckling. Similar to MXene metamaterials with straight ligaments, the out-of-plane deformations induced by ligament buckling under compression tend to intensify the auxetic response.

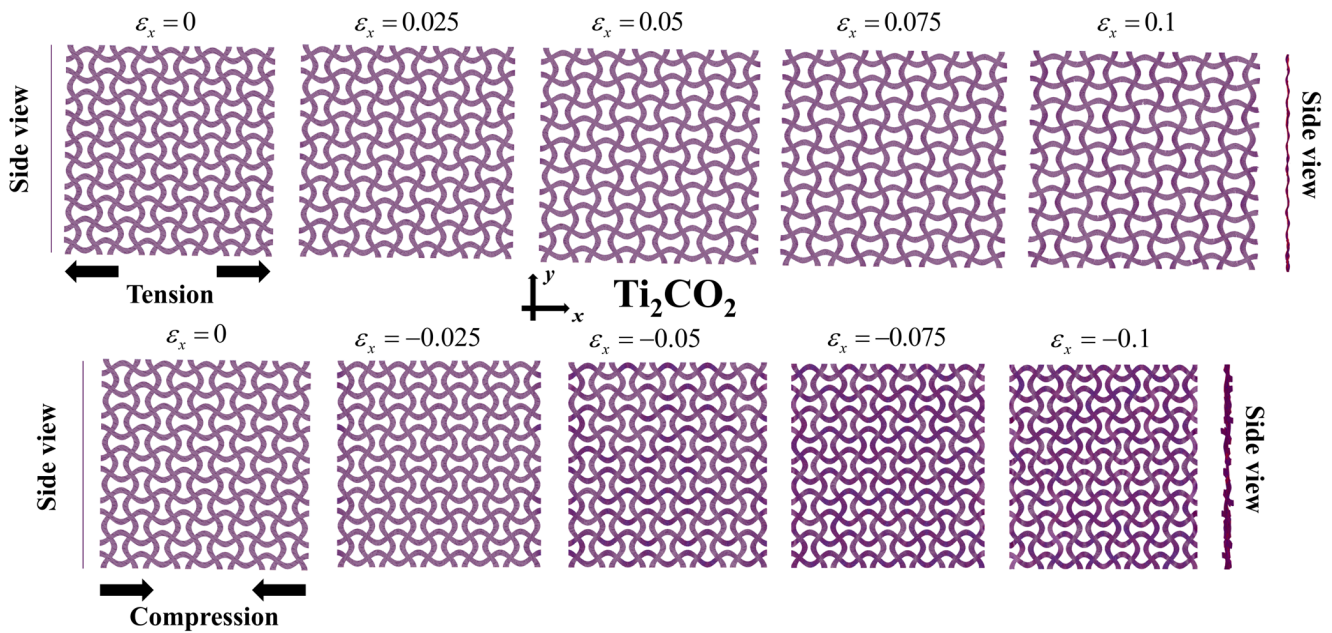

**Fig. 12** Atomic configurations (top view) of the $Ti_2CO_2$ MXene metamaterial with curved ligaments subjected to uniaxial tension and compression along the zigzag direction. Results are shown for a temperature of 1 K, $t/L=0.125$, $l/d=3$, and different strain levels. Side views of the initial and last snapshots are also shown.

Movies S7 and S8 (tension) and S9 and S10 (compression), provided in the Supplementary Materials, illustrate the complex deformation processes in $Ti_2C$ and $Ti_2CO_2$ MXene metamaterials with curved ligaments. In general, the magnitude of out-of-plane deformation is smaller in MXene metamaterials with oxygen termination than in pristine MXenes. This behavior arises because surface termination effectively increases the sheet thickness and, consequently, enhances the bending rigidity. In contrast, more pronounced out-of-plane deformations occur in the absence of surface termination and in structures with thinner ligaments.

The in-plane and out-of-plane deformations of the central unit cell of a $Ti_2C$ MXene metamaterial with curved ligaments subjected to uniaxial tension along the zigzag direction are shown in Fig. 13 for different strain levels. This figure illustrates the evolution of the deformation mechanisms during loading. The out-of-plane deformation increases progressively with increasing strain. At higher strain levels, the ligaments not only deform out of the plane but also undergo complex twisting. This auxetic out-of-plane response is a distinctive feature of 2D metamaterials. The in-plane deformation is primarily characterized by rotations of the junctions, which drive the lateral expansion of the structure. In addition, the ligaments aligned with the loading direction tend to flatten more than those oriented perpendicular to the loading direction.

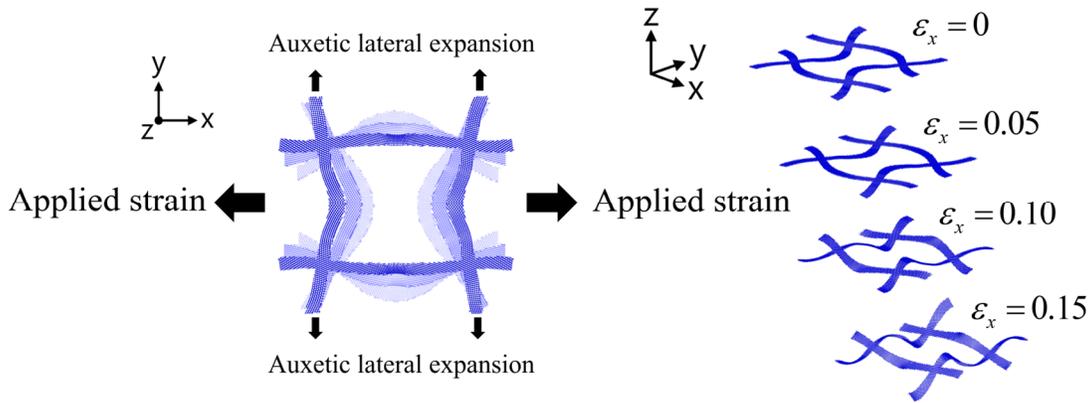

**Fig. 13** In-plane and out-of-plane deformations of the central unit cell of $Ti_2C$ MXene metamaterial with curved ligaments subjected to uniaxial tension along the zigzag direction. Results are shown for a temperature of 1 K, $t/L=0.05$, $l/d=3$, and different strain levels.

To further investigate the mechanical behavior of MXene metamaterials, we analyze the evolution of stress distributions within these structures. Atomic stresses calculated from the virial theorem are typically noisy at the atomistic scale. To improve the resolution of the stress fields, a local spatial averaging procedure is employed. This approach, which is an efficient tool for obtaining continuum-like stress distributions even in regions with high stress gradients using atomistic data [63–65], assumes a spherical neighborhood around each atom with a prescribed radius. The center of this sphere coincides with the atom of interest. In the present study, a radius of 5 Å is adopted. All neighboring atoms located within this sphere are used to compute a spatial average of each stress component, and the resulting averaged values are assigned to the central atom.

Using this local spatial averaging technique, the evolution of the in-plane shear stress within the central unit cells of a $Ti_2CO_2$ MXene metamaterial with straight ligaments and a $Ti_2C$ MXene metamaterial with curved ligaments is presented in Fig. 14 for increasing strain levels. The simulations are performed at a temperature of 1 K. The stress distributions are displayed in a dimensionless form.

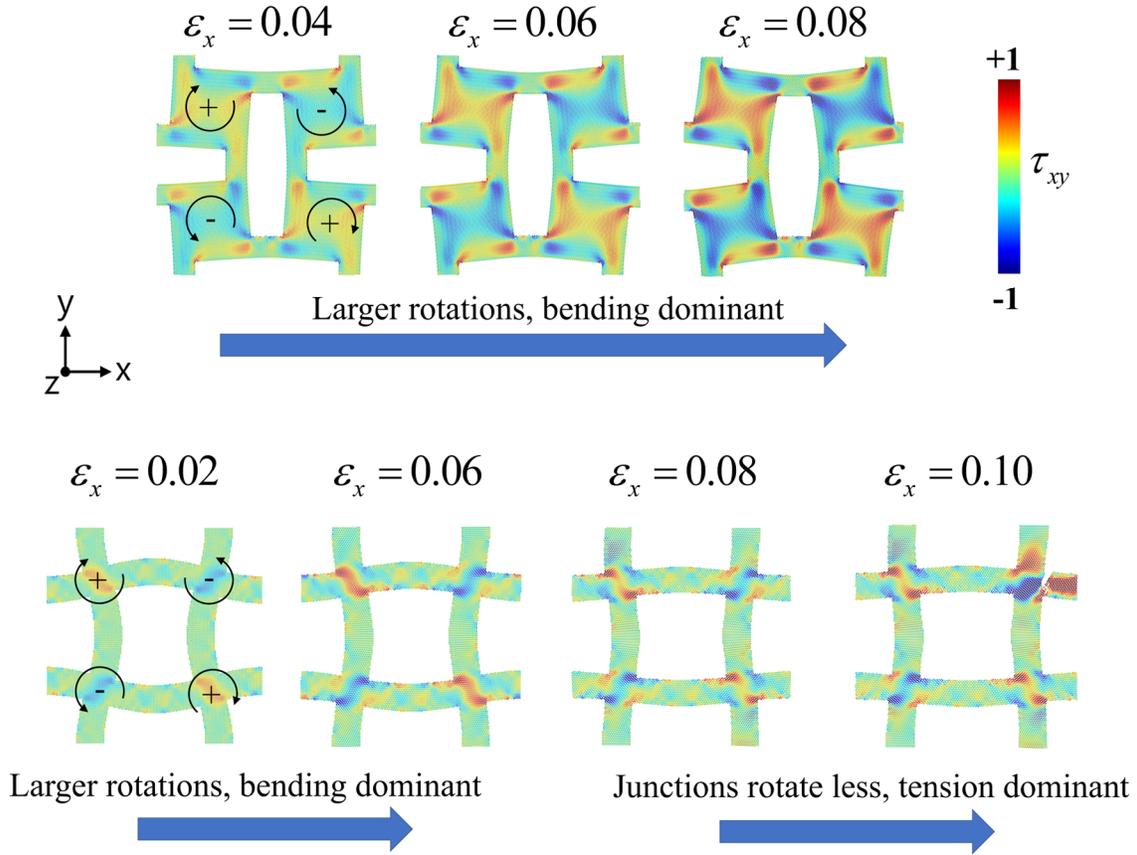

**Fig. 14** Evolution of the in-plane shear stress in the unit cells at the center of a $Ti_2CO_2$ MXene metamaterial with straight ligaments ($t/L=0.1$, $l/d=5$), and a $Ti_2C$ MXene metamaterial unit cell with curved ligaments ($t/L=0.125$, $l/d=3/2$) on increasing strain level. Temperature is 1 K. The stress distributions are shown in a dimensionless form.

In both perforation geometries, the shear stresses reach their maximum values near the corners of the perforations, exhibit an alternating pattern around the perforation, with two diagonally opposite junctions experiencing positive shear and the other two experiencing negative shear, and gradually decay along the ligaments. In the MXene metamaterial with rectangular perforations, the shear stress increases continuously as the applied strain rises from 0.04 to 0.08. Progressively larger rotations of the junctions accompany this increase. In contrast, for the metamaterial with curved ligaments, the shear stress initially increases up to a strain of 0.06, with the junction rotations also increasing, indicating that bending dominates the deformation mechanism in this regime. However, as the strain increases from 0.06 to 0.08, the shear stress becomes more concentrated at the junction corners, and a transition to a tension-dominated deformation mechanism occurs. At a strain of 0.1, a crack initiates from the corner of the perforation.

Similarly, the evolution of axial stresses is shown in Fig. 15. In the MXene metamaterials with rectangular perforations, the maximum tensile stresses occur at the corners of the perforations due to stress concentrations. For metamaterials with curved ligaments, the maximum tensile stress is observed at the inner regions of the curved ligaments aligned with the loading direction, typically at the midspan of the ligament. In this case, each ligament behaves like a curved beam that bends to flatten under tension. In both geometries, cracks initiate in the regions of maximum tensile stress. Comparing the curved-ligament metamaterials in Fig. 14 and Fig. 15, it is evident that surface termination and perforation geometry influence the sites of crack initiation. In general, cracks appear either at the perforation corners, where stress concentrations are high, or at the midpoint of the curved ligaments. The complex 3D deformation of the unit cells, particularly the one with a rectangular perforation, is evident in Fig. 15.

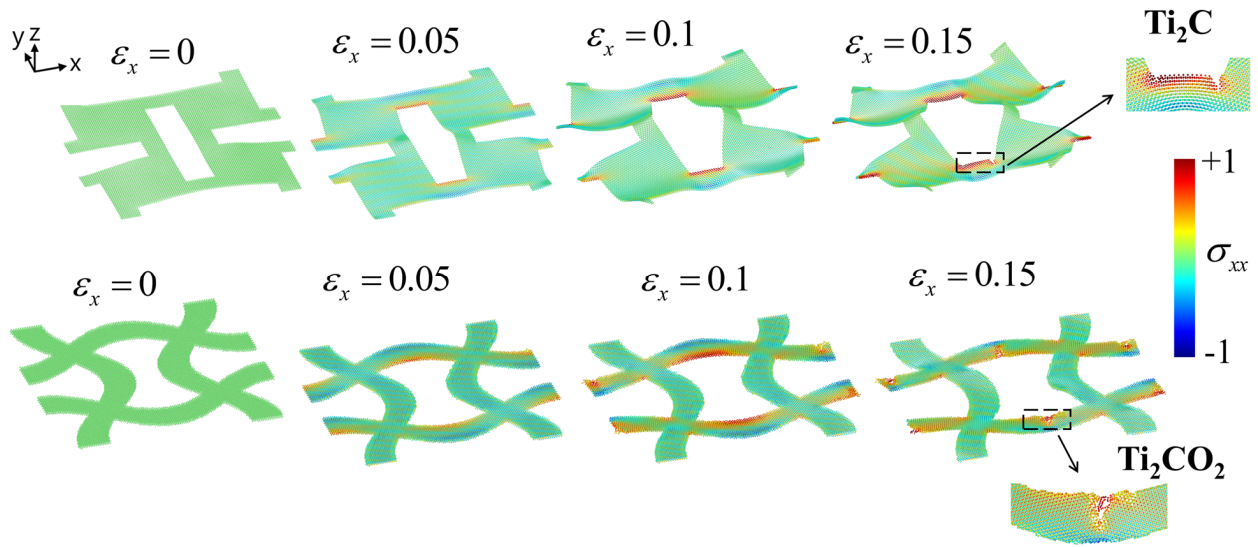

**Fig. 15** Evolution of the uniaxial tensile stress along the *x*-axis in the unit cells at the center of a $Ti_2C$ MXene metamaterial with straight ligaments ($t/L=0.1$, $l/d=5$), and a $Ti_2CO_2$ MXene metamaterial with curved ligaments ($t/L=0.125$, $l/d=3$) on increasing strain level. Temperature is 1 K. The stress distributions are shown in a dimensionless form.

Atomic configurations and in-plane shear stress distributions in the central unit cells of $Ti_2CO_2$ MXene metamaterials with straight and curved ligaments subjected to uniaxial compression along the zigzag direction are shown in Fig. 16. Under compression, the perforations tend to shrink due to rotations of the junctions and the applied compressive strain. As evident in Fig. 16(c) and (d), the shear stress distributions are similar to those shown in Fig. 14 for the tensile case; however, their signs are reversed, indicating that the in-plane rotations of the junctions occur in the opposite direction relative to those under tension. These rotations, coupled with the applied compressive forces, cause the ligaments aligned with the loading direction to buckle, leading to complex 3D deformations within the unit cell. In both perforation geometries, shear stress at the junctions increases with applied strain, indicating greater junction rotation at higher strain levels.

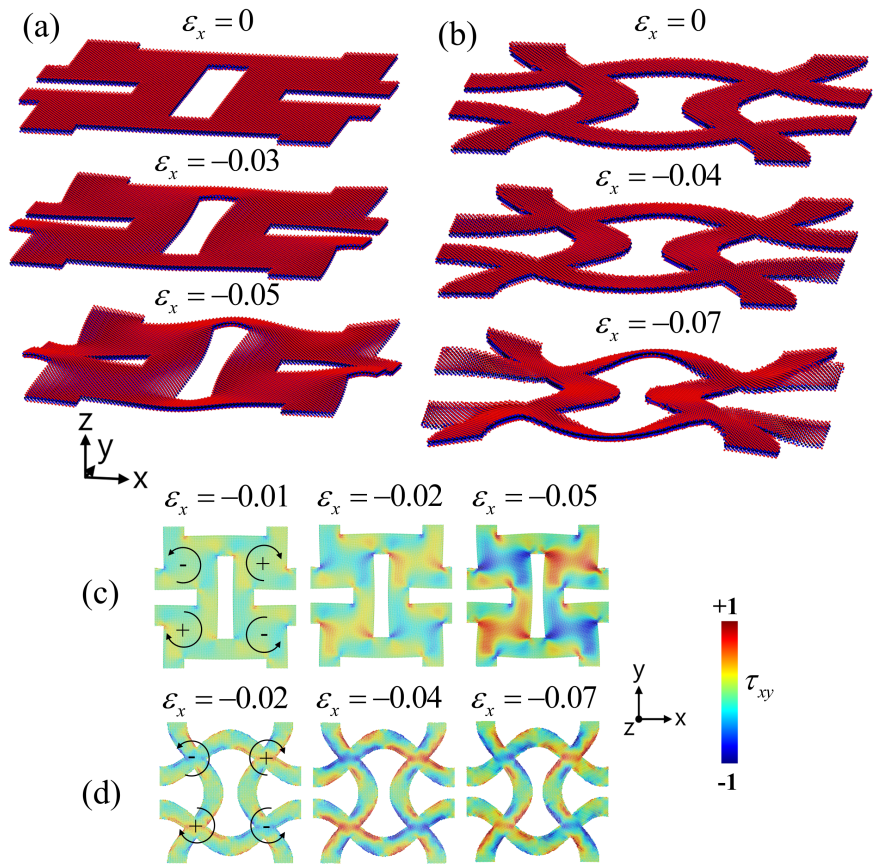

**Fig. 16** (a) and (b) Atomic configurations, and (c) and (d) in-plane shear stresses of the unit cells at the center of $Ti_2CO_2$ MXene metamaterials with straight ($t/L=0.15$, $l/d=5$, ) and curved ($t/L=0.125$, $l/d=3$) ligaments subjected to uniaxial compression along the zigzag direction. Results are presented at a temperature of 1 K and for different strain levels. The stress distributions are shown in a dimensionless form.

## 3.5. Temperature Effect

All results presented so far correspond to simulations performed at 1 K. To investigate the effect of temperature, additional tensile simulations are conducted on a $Ti_2C$ MXene metamaterial with straight ligaments at three higher temperatures: 150 K, 300 K, and 450 K. The corresponding stress–strain, axial strain–lateral strain, and axial strain–Poisson's ratio curves are shown in Fig. 17.

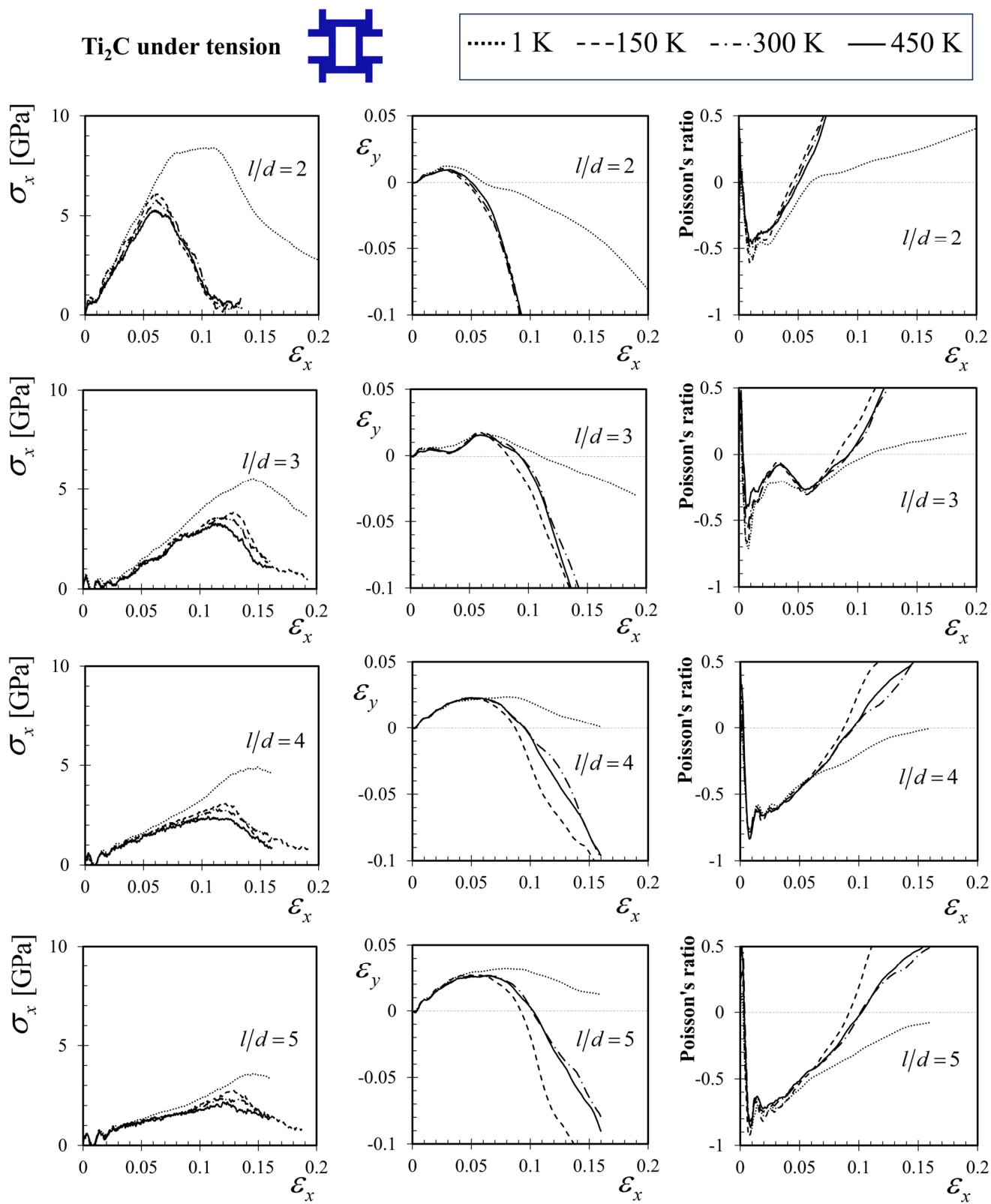

**Fig. 17** Stress–strain response, axial and lateral strain evolution, and Poisson's ratio curves of the Ti$_2$C MXene metamaterial with straight ligaments ($t/L=0.10$) subjected to uniaxial tension along the zigzag direction. Results are shown for different temperatures and rectangular perforation aspect ratios, $l/d$.

The initial stiffness of the stress–strain curves decreases slightly with increasing temperature, while the ultimate tensile strength of the MXene metamaterial declines significantly at higher temperatures. The lateral strain and, consequently, Poisson's ratio appear to be much less sensitive to temperature, particularly at lower strain levels. This observation is in agreement with the data reported in Ref. [66] for graphene metamaterials, and emphasizes the practical relevance of the 1 K findings presented in this paper for the design of MXene metamaterials across a wide range of temperatures.

Nevertheless, the auxetic response at higher strains is slightly reduced as the temperature increases. One possible explanation is that higher temperatures lead to more pronounced out-of-plane deformations due to increased atomic thermal vibrations and reduced bending rigidity. As discussed earlier, increased out-of-plane deformations under tension tend to diminish the auxetic response. This effect is illustrated in Movie S11 provided in the Supplementary Materials.

### 3.6. Ti$_3$C$_2$O$_2$ MXene Metamaterial

The mechanical response of Ti$_3$C$_2$O$_2$ MXene metamaterials with straight ligaments is analyzed in this section. Due to the large number of atoms involved in the simulations, we focus on a single ligament thickness of $t/L=0.10$ and a temperature of 1 K. The corresponding stress–strain, axial strain–lateral strain, and axial strain–Poisson's ratio curves are shown in Fig. 18(a), while the out-of-plane deformations and atomic axial and shear stress distributions are illustrated in Fig. 18(b), (c), and (d) at increasing applied strain. Comparing the stress–strain curves in Fig. 18(a) with those of Ti$_2$CO$_2$ in Fig. 3 reveals that Ti$_3$C$_2$O$_2$ exhibits a higher tensile strength. The auxetic response of both MXene metamaterials is similar, with higher perforation aspect ratios producing a more pronounced NPR. Out-of-plane deformations in Fig. 18(b) are smaller than those observed for Ti$_2$CO$_2$ in Fig. 11(d), reflecting the higher bending rigidity of Ti$_3$C$_2$O$_2$ due to its larger effective thickness. The axial and shear stress distributions show patterns consistent with those presented for Ti$_2$C and Ti$_2$CO$_2$ in Fig. 14 and Fig. 15, with maximum axial stress occurring at the corners of the perforations, where cracks initiate at a strain of 0.088. Junction rotations are also evident, confirming that the "rotating rigid units" deformation mode is the primary mechanism driving the auxetic response in Ti$_3$C$_2$O$_2$ MXene metamaterials.

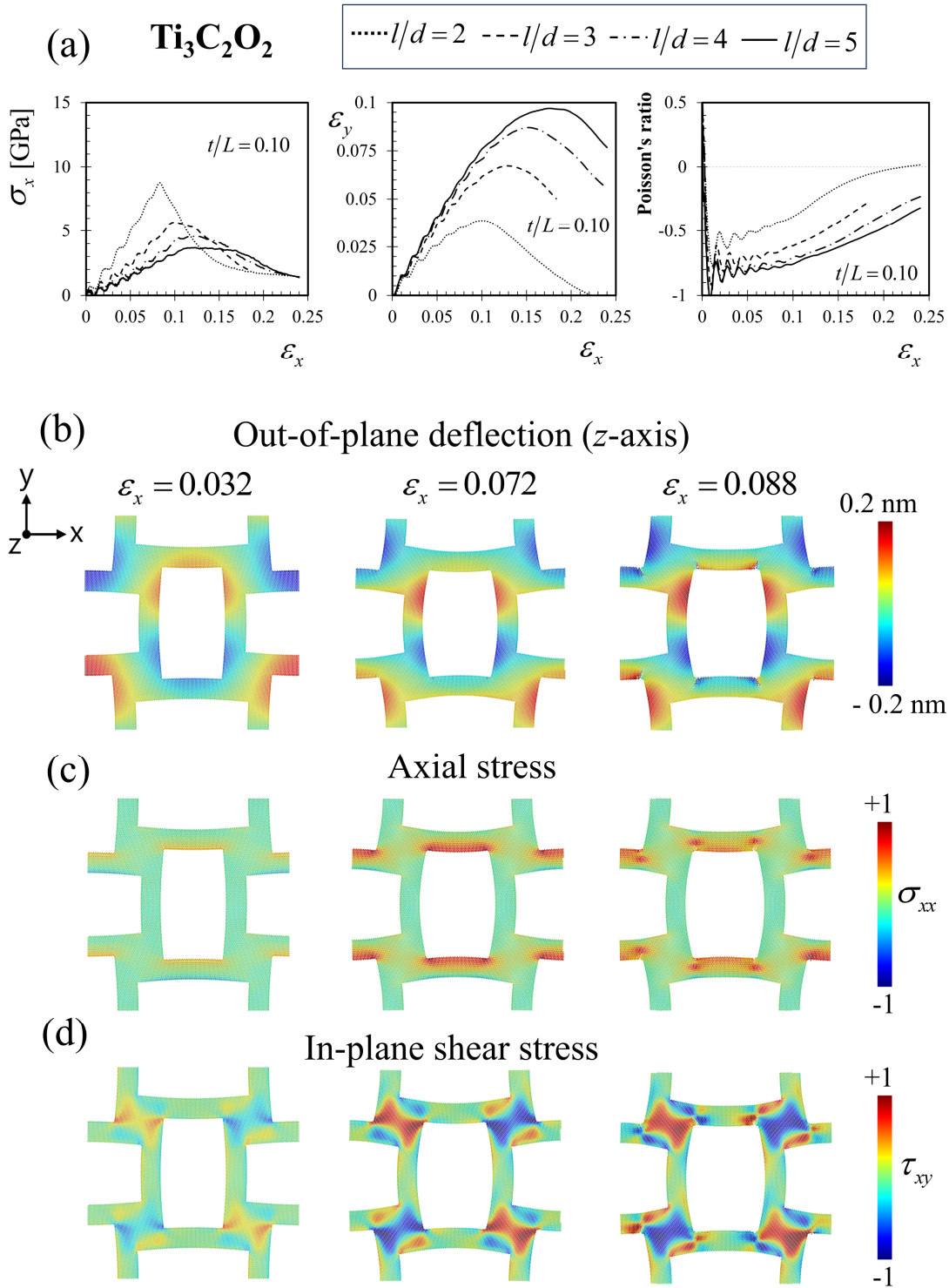

**Fig. 18** (a) Stress–strain response, axial and lateral strain evolution, and Poisson's ratio curves, (b) out-of-plane deflection, (c) axial stress, and (d) in-plane shear stress distributions in the $Ti_3C_2O_2$ MXene metamaterial with straight ligaments ($t/L=0.10$) subjected to a uniaxial tension along the zigzag direction at 1 K. The deflection and stress distributions are presented for $l/d=2$. Cracks at the corners of the perforation are visible at a strain of 0.088.

## 3.7. MXene vs. Graphene Metamaterials

Tables 2 and 3 summarize the Poisson's ratios obtained for some of the studied MXene metamaterials with straight and sinusoidally curved ligaments, respectively. The values for MXene metamaterials are obtained from a linear fit of the axial–lateral strain curves within the strain range of 0–2%. The results are reported for both $Ti_2C$ and $Ti_2CO_2$ MXenes and compared with previously reported values for graphene metamaterials with the same geometric configurations using MD simulations and the AIREBO interatomic potential (Refs. [11,66]). This comparison enables an assessment of the extent to which the auxetic response depends on the intrinsic properties of the underlying 2D material.

**Table 2:** Poisson's ratios of MXene and graphene metamaterials with straight ligaments for different perforation thicknesses and aspect ratios. The values for MXene metamaterials are obtained from a linear fit of the axial–lateral strain curves within the strain range of 0–2%. The graphene values are taken from the Supplementary Materials of Ref. [66].

| t/L | l/d | Poisson's Ratio $Ti_2C$ MXene / Poisson's Ratio $Ti_2CO_2$ MXene / Poisson's Ratio Graphene [66] |
|---|---|---|
| 0.25 | 2 | -0.04 |
|  |  | 0.13 |
|  |  | 0.12 |
| 0.1 | 2 | -0.48 |
|  |  | -0.54 |
|  |  | -0.51 |
| 0.25 | 4 | -0.13 |
|  |  | -0.01 |
|  |  | -0.02 |
| 0.2 | 2 | -0.09 |
|  |  | -0.02 |
|  |  | -0.05 |
| 0.2 | 4 | -0.25 |
|  |  | -0.27 |
|  |  | -0.21 |
| 0.15 | 2 | -0.26 |
|  |  | -0.26 |
|  |  | -0.28 |
| 0.1 | 4 | -0.63 |
|  |  | -0.77 |
|  |  | -0.83 |
| 0.1 | 5 | -0.75 |
|  |  | -0.79 |
|  |  | -0.89 |
| 0.15 | 5 | -0.57 |
|  |  | -0.59 |
|      |   | -0.70 |
|      |   | -0.29 |
| 0.2  | 5 | -0.32 |
|      |   | -0.24 |
|      |   | -0.16 |
| 0.25 | 5 | -0.05 |
|      |   | 0.06  |

**Table 3:** Poisson's ratios of MXene and graphene metamaterials with sinusoidally curved ligaments for different perforation thicknesses and aspect ratios. The values for MXene metamaterials are obtained from a linear fit of the axial–lateral strain curves within the strain range of 0–2%. The graphene values are taken from the curves in Ref. [11].

| $t/L$ | $l/d$ | Poisson's Ratio $Ti_2C$ MXene<br>Poisson's Ratio $Ti_2CO_2$ MXene<br>Poisson's Ratio Graphene [11] |
|-------|-------|---|
| 0.05  | 3     | -0.63 |
|       |       | -0.57 |
|       |       | -0.67 |
| 0.075 | 3     | -0.61 |
|       |       | -0.64 |
|       |       | -0.54 |
| 0.010 | 3     | -0.53 |
|       |       | -0.57 |
|       |       | -0.46 |
| 0.125 | 3     | -0.45 |
|       |       | -0.48 |
|       |       | -0.41 |
| 0.05  | 3/2   | -0.27 |
|       |       | -0.31 |
|       |       | -0.36 |
| 0.05  | 13/7  | -0.46 |
|       |       | -0.52 |
|       |       | -0.50 |
| 0.05  | 7/3   | -0.59 |
|       |       | -0.59 |
|       |       | -0.58 |

Although the numerical values of the Poisson ratio differ among the three material systems, all exhibit a clear auxetic response over the examined range of geometrical parameters, except for two cases corresponding to rectangular perforations with a dimensionless ligament thickness of 0.25 and aspect ratios of 2 and 5.

A comparison of the results shows that the geometric characteristics of the perforated lattice influence the auxetic response in a similar manner in both MXene and graphene metamaterials. In particular, increasing the perforation aspect ratio increases NPR in both systems. Likewise, increasing the ligament thickness generally weakens the auxetic response. These trends indicate that the geometry of the perforated architecture plays a dominant role in the auxetic behavior of the 2D metamaterial.

Despite overall similarity in deformation mechanisms, NPR differs due to the distinct mechanical properties of MXenes and graphene. MXenes possess a larger effective thickness and different bending rigidity compared with graphene, which leads to variations in the resulting Poisson's ratios even for identical geometries. For example, a graphene metamaterial with straight ligaments characterized by $t/L = 0.25$ and $l/d = 2$ exhibits a positive Poisson ratio of 0.12. A $Ti_2CO_2$ metamaterial with the same geometry shows a very similar response with a Poisson ratio of 0.13. In contrast, the $Ti_2C$ metamaterial with the same geometry exhibits a negative Poisson ratio of −0.04.

These observations indicate that while the geometric design primarily governs the emergence of auxetic behavior, the intrinsic mechanical properties of the underlying 2D material influence the magnitude of the response. This combination of geometry-driven mechanisms and material-dependent characteristics provides additional flexibility for tailoring the mechanical performance of such metamaterials, allowing designers to select both the architecture and the base material to achieve targeted auxetic responses and mechanical properties.

## 4. Conclusions

In summary, this study demonstrates that perforated MXenes, which have previously been explored primarily for non-mechanical applications such as filtration and molecular separation, can exhibit pronounced auxetic mechanical behavior. This finding is established through large-scale reactive molecular dynamics simulations of tensile and compressive deformation in MXene monolayers of $Ti_2C$, $Ti_2CO_2$, and $Ti_3C_2O_2$, containing engineered perforations with straight or sinusoidally curved ligaments. The effects of perforation geometry, ligament thickness, surface termination, loading mode, and temperature were systematically investigated.

The results show that the auxetic response originates from a rotating-junction mechanism driven by alternating in-plane shear stresses localized at the ligament junctions. This mechanism is intrinsically coupled to out-of-plane deformation arising from the finite bending rigidity of atomically thin sheets, producing complex 3D distortions even under in-plane loading. Increasing the perforation aspect ratio intensifies the auxetic response but reduces stiffness and ultimate strength, whereas increasing ligament thickness enhances stiffness and strength while weakening the negative Poisson ratio. Surface termination alters the effective thickness and bending rigidity of MXenes, resulting in quantitative differences in the auxetic response while preserving the underlying deformation mechanism. Temperature is found to reduce stiffness and strength moderately, but only slightly diminishes the auxetic behavior. Comparison with graphene metamaterials further reveals that the qualitative auxetic trends in 2D materials are primarily governed by geometry, whereas quantitative differences arise from intrinsic material properties such as bending rigidity.

This paper highlights the potential of MXenes as a versatile candidate for auxetic 2D materials. Owing to the large compositional diversity of the MXene family and the tunability of their surface chemistry, perforated MXene sheets offer significantly greater flexibility for the design of auxetic nanostructures compared with other 2D materials such as graphene.

## Declaration of interests

The author declares that he has no known competing financial interests or personal relationships that could have appeared to influence the work reported in this paper.

## Funding

This research did not receive any specific grant from funding agencies in the public, commercial, or not-for-profit sectors.

## Acknowledgements

The author gratefully acknowledges the Polish high-performance computing infrastructure PLGrid (HPC Centers: WCSS, ACK Cyfronet AGH) for providing computer facilities and support within the computational grant no. PLG/2026/019291.